\documentclass[traditabstract,longauth,oldversion]{aa} 

\usepackage{graphicx}
\usepackage{txfonts}
\usepackage{natbib}
\usepackage[usenames,dvipsnames]{color}
\bibpunct{(}{)}{;}{a}{}{,} 

\newcommand{\psr}{PSR\,B1706$-$44}
\newcommand{\snr}{G\,343.1$-$2.3}
\newcommand{\hessj}{HESS\,J1708$-$443}

\begin{document}

\title{Detection of very-high-energy $\gamma$-ray emission from the vicinity of \psr\ and \snr\ with H.E.S.S.}

\author{ H.E.S.S.\ Collaboration:
 \and A.~Abramowski \inst{4}
 \and F.~Acero \inst{15}
 \and F. Aharonian\inst{1,13,35}
 \and A.G.~Akhperjanian \inst{2,35}
 \and G.~Anton \inst{16}
 \and A.~Barnacka \inst{24,7}
 \and U.~Barres de Almeida \inst{8} \thanks{supported by CAPES Foundation, Ministry of Education of Brazil}
 \and A.R.~Bazer-Bachi \inst{3}
 \and Y.~Becherini \inst{12,10}
% ...
}

\institute{
Max-Planck-Institut f\"ur Kernphysik, P.O. Box 103980, D 69029
Heidelberg, Germany
\and
 Yerevan Physics Institute, 2 Alikhanian Brothers St., 375036 Yerevan,
Armenia
\and
Centre d'Etude Spatiale des Rayonnements, CNRS/UPS, 9 av. du Colonel Roche, BP
4346, F-31029 Toulouse Cedex 4, France
\and
Universit\"at Hamburg, Institut f\"ur Experimentalphysik, Luruper Chaussee
149, D 22761 Hamburg, Germany
\and
Institut f\"ur Physik, Humboldt-Universit\"at zu Berlin, Newtonstr. 15,
D 12489 Berlin, Germany
\and
LUTH, Observatoire de Paris, CNRS, Universit\'e Paris Diderot, 5 Place Jules Janssen, 92190 Meudon, 
France
\and
CEA Saclay, DSM/IRFU, F-91191 Gif-Sur-Yvette Cedex, France
\and
University of Durham, Department of Physics, South Road, Durham DH1 3LE,
U.K.
\and
Unit for Space Physics, North-West University, Potchefstroom 2520,
    South Africa
\and
Laboratoire Leprince-Ringuet, Ecole Polytechnique, CNRS/IN2P3,
 F-91128 Palaiseau, France
\and 
Laboratoire d'Annecy-le-Vieux de Physique des Particules,
Universit\'{e} de Savoie, CNRS/IN2P3, F-74941 Annecy-le-Vieux,
France
\and
Astroparticule et Cosmologie (APC), CNRS, Universit\'{e} Paris 7 Denis Diderot,
10, rue Alice Domon et L\'{e}onie Duquet, F-75205 Paris Cedex 13, France
\thanks{UMR 7164 (CNRS, Universit\'e Paris VII, CEA, Observatoire de Paris)}
\and
Dublin Institute for Advanced Studies, 5 Merrion Square, Dublin 2,
Ireland
\and
Landessternwarte, Universit\"at Heidelberg, K\"onigstuhl, D 69117 Heidelberg, Germany
\and
Laboratoire de Physique Th\'eorique et Astroparticules, 
Universit\'e Montpellier 2, CNRS/IN2P3, CC 70, Place Eug\`ene Bataillon, F-34095
Montpellier Cedex 5, France
\and
Universit\"at Erlangen-N\"urnberg, Physikalisches Institut, Erwin-Rommel-Str. 1,
D 91058 Erlangen, Germany
\and
Laboratoire d'Astrophysique de Grenoble, INSU/CNRS, Universit\'e Joseph Fourier, BP
53, F-38041 Grenoble Cedex 9, France 
\and
Institut f\"ur Astronomie und Astrophysik, Universit\"at T\"ubingen, 
Sand 1, D 72076 T\"ubingen, Germany
\and
LPNHE, Universit\'e Pierre et Marie Curie Paris 6, Universit\'e Denis Diderot
Paris 7, CNRS/IN2P3, 4 Place Jussieu, F-75252, Paris Cedex 5, France
\and
Charles University, Faculty of Mathematics and Physics, Institute of 
Particle and Nuclear Physics, V Hole\v{s}ovi\v{c}k\'{a}ch 2, 
180 00 Prague 8, Czech Republic
\and
Institut f\"ur Theoretische Physik, Lehrstuhl IV: Weltraum und
Astrophysik,
    Ruhr-Universit\"at Bochum, D 44780 Bochum, Germany
\and
University of Namibia, Department of Physics, Private Bag 13301, Windhoek, Namibia
\and
Obserwatorium Astronomiczne, Uniwersytet Jagiello{\'n}ski, ul. Orla 171,
30-244 Krak{\'o}w, Poland
\and
Nicolaus Copernicus Astronomical Center, ul. Bartycka 18, 00-716 Warsaw,
Poland
 \and
School of Physics \& Astronomy, University of Leeds, Leeds LS2 9JT, UK
 \and
School of Chemistry \& Physics,
 University of Adelaide, Adelaide 5005, Australia
 \and 
Toru{\'n} Centre for Astronomy, Nicolaus Copernicus University, ul.
Gagarina 11, 87-100 Toru{\'n}, Poland
\and
Instytut Fizyki J\c{a}drowej PAN, ul. Radzikowskiego 152, 31-342 Krak{\'o}w,
Poland
\and
Astronomical Observatory, The University of Warsaw, Al. Ujazdowskie
4, 00-478 Warsaw, Poland
\and
Institut f\"ur Astro- und Teilchenphysik, Leopold-Franzens-Universit\"at 
Innsbruck, A-6020 Innsbruck, Austria
\and
Oskar Klein Centre, Department of Physics, Stockholm University,
Albanova University Center, SE-10691 Stockholm, Sweden
\and
Oskar Klein Centre, Department of Physics, Royal Institute of Technology (KTH),
Albanova, SE-10691 Stockholm, Sweden
\and
Department of Physics and Astronomy, The University of Leicester, 
University Road, Leicester, LE1 7RH, United Kingdom
\and
European Associated Laboratory for Gamma-Ray Astronomy, jointly
supported by CNRS and MPG
\and
National Academy of Sciences of the Republic of Armenia, Yerevan 
% ...
}

\offprints{ryan.chaves@mpi-hd.mpg.de, s.d.hoppe@googlemail.com}

\date{Received 13 July 2010; Accepted 28 November 2010}

\abstract{The $\gamma$-ray pulsar \object{PSR\,B1706$-$44} and the
  adjacent supernova remnant (SNR) candidate \object{G\,343.1$-$2.3} were observed by
  H.E.S.S.\ during a dedicated observation campaign in 2007.
  As a result of this observation campaign, a new source of very-high-energy (VHE; E~$>$~100~GeV) $\gamma$-ray
  emission, \object{\hessj}, was detected with a statistical significance
  of 7 $\sigma$, although no significant point-like emission was detected at the position of the energetic pulsar itself.
  In this paper, the morphological and spectral analyses of the
  newly-discovered TeV source are presented. 
  The centroid of \hessj\ is considerably offset from the pulsar 
  and located near the apparent center of the SNR, at
  $\alpha_{\mathrm{J2000}} = 17^{\mathrm{h}}08^{\mathrm{m}}11^{\mathrm{s}} \pm 17^{\mathrm{s}}$ and
  $\delta_{\mathrm{J2000}} = -44\degr20\arcmin \pm 4\arcmin$.
  The source is found to be significantly more
  extended than the H.E.S.S.\ point spread function ($\sim$0.1\degr), with an intrinsic Gaussian width
  of 0.29\degr~$\pm$~0.04$\degr$.
  Its integral flux
  between 1 and 10~TeV is $\sim3.8 \times 10^{-12}$~ph~cm$^{-2}$~s$^{-1}$, equivalent to 17\% of the \object{Crab~Nebula} 
  flux in the
  same energy range. The measured energy spectrum is well-fit by
  a power law with a relatively hard photon index
  $\Gamma$~=~2.0~$\pm$~0.1$_{\mathrm{stat}}$~$\pm$~$0.2_{\mathrm{sys}}$. 
  Additional multi-wavelength data, including 330~MHz VLA observations, were used
  to investigate the VHE $\gamma$-ray source's possible associations with the pulsar wind nebula of
  \psr\ and/or with the complex 
  radio structure of the partial shell-type SNR \snr.
}


\authorrunning{H.E.S.S. Collaboration}
\titlerunning{VHE $\gamma$-rays from \psr\ \& SNR \snr}
\maketitle

\section{Introduction}
The energetic pulsar \psr\ (also \object{PSR\,J1709$-$4429}) is one of
the first pulsars from which pulsed emission was detected not only in
the radio \citep{Johnston_1992} and in X-rays \citep{Gotthelf_2002}, but also
in high-energy (HE; E~$\sim$~GeV) $\gamma$-rays \citep{Swanenburg_1981}.  The pulsar
was first detected in a high-frequency radio
survey by \cite{Johnston_1992} and has a spin period $P$~$=$~102~ms, a
characteristic age $\tau_c$~$=$~17\,500 yr, and a spin-down luminosity
$\dot{E} = 3.4 \times 10^{36} \mathrm{erg} \mathrm{s}^{-1}$.
It belongs to the class of relatively young and powerful pulsars, 
of which the \object{Vela Pulsar} is the most prominent example in the southern hemisphere.  
The putative wind nebulae of these
pulsars are prime candidates for being sources of very-high-energy (VHE; E~$>$~100~GeV) $\gamma$-rays.
A bright, HE $\gamma$-ray source,
\object{2CG\,342$-$02}, was discovered by the \emph{COS-B} satellite \citep{Swanenburg_1981}
and later found to be positionally coincident with the radio pulsar.  The $\gamma$-ray source was firmly associated with 
\psr\ after EGRET (the Energetic Gamma Ray Experiment Telescope, onboard 
the \emph{Compton Gamma-Ray Observatory}) observed pulsations from \object{3EG\,J1710$-$4439} 
(also \object{EGR\,J1710$-$4435}) which matched the period seen
in the radio waveband \citep{Thompson_1992}.  More recently, the pulsar has been detected at GeV energies  
by the latest generation of spaceborne HE $\gamma$-ray detectors: 
by \emph{AGILE} (\emph{Astrorivelatore Gamma ad Immagini LEggero})
as \object{1AGL\,J1709$-$4428} \citep{Pittori2009} and by the
\emph{Fermi}/LAT (Large Area Telescope) as \object{1FGL\,J1709.7$-$4429} \citep{Abdo2010LAT1FGL}.

Radio observations of \psr\ 
reveal the presence of a synchrotron nebula, with an extension of $\sim$3\arcmin,
surrounding the
pulsar \citep{Frail_1994,Giacani_2001}. The observed polarization and the flat spectrum, 
with a flux density spectral slope $\alpha = 0.3$ (where the flux density $S \propto \nu^{-\alpha}$),
suggest it is a pulsar wind nebula (PWN). However, the
implied conversion efficiency from spin-down energy to radio flux of
$\sim$2~$\times$~10$^{-6}$ would be the lowest of any known radio PWN
\citep{Giacani_2001}. Observations by the X-ray telescopes onboard \emph{ROSAT} (\emph{Roentgen Satellite})
and \emph{ASCA} (\emph{Advanced Satellite for Cosmology and Astrophysics})
reveal that the 
nebula is also visible in X-rays \citep{Finley_1998}.
The morphology of the PWN was mapped in detail at arcsecond scales utilizing the superior resolution of the
\emph{Chandra X-ray Observatory} \citep{Romani_2005}.
The X-ray analyses suggest the presence of a diffuse X-ray PWN, with a 
spectral index of 1.77, which surrounds a more
complex structure consisting of a torus and inner and outer jets.
The diffuse X-ray PWN has a radius of 1.8\arcmin~and also exhibits a fainter, longer extension to the West.
The presence of
non-deformed X-ray jets is consistent with the pulsar's low apparent speed, $v = 89$~km~s$^{-1}$, as deduced from
scintillation measurements \citep{Johnston_1998}.

The pulsar \psr\ is also located at the southeast end of an incomplete arc
of radio emission \citep{McAdam_1993}, which has been suggested to be the partial shell of a
faint supernova remnant (SNR \snr). The arc is embedded in weak
diffuse radio emission, which is present both inside and outside of the arc \citep{Frail_1994}.
Polarization measurements suggest that this diffuse emission is associated with
synchrotron radiation from the SNR itself \citep{Dodson_2002}. The SNR has
not been detected in any other waveband \citep[see
e.g.][]{Becker_1995,Aharonian_2005_1706}.
There are various estimates of the distance to the pulsar, 
ranging from 1.8~kpc \citep{Johnston_1992,Taylor_1993} to 3.2~kpc \citep{Koribalski_1995}.
The distance 2.3~$\pm$~0.3~kpc, derived from the dispersion measure and the most recent Galactic free
electron distribution model \citep{Cordes_2002}, is adopted throughout this paper.
This distance is compatible with the less
reliable $\Sigma-D$ distance of $\sim$3~kpc for the SNR
\citep{McAdam_1993}.

The possible physical association between
\psr\ and \snr\ has been questioned based on the
differing age and distance estimates for the SNR and pulsar (see Sec.\ 4.2 and 4.3, respectively) and
the lack of visible interaction.  Furthermore, if the pulsar originated at the apparent center of the SNR, then
its inferred velocity ($\sim$700~km~s$^{-1}$) is incompatible with its scintillation velocity (89~km~s$^{-1}$).
\cite{Bock_2002} suggested a scenario where an
off-center cavity explosion could relax the restrictions on
the inferred velocity and invalidate the age estimate for the SNR of
$\sim$5\,000 yr \citep{McAdam_1993}, which is based on a Sedov-Taylor
model. In this scenario, \psr\ and \snr\ are physically
associated; however, the radio arc is not identified with the SNR
shell, but rather with the former boundary of the wind-blown cavity that was
overtaken and compressed by the expanding SNR \citep{Dodson_2002}.

\begin{table*}[]
  \begin{center}
    \begin{tabular}{c c c c c c}
      \hline\hline
      Observation dates & Instrument    & Test position & Extension & Integral flux (ph cm$^{-2}$ s$^{-1}$)                     & Reference \\ 
      \hline
      1992              & CANGAROO-I    & \psr        & n/a       & $F(> 1\,\mathrm{TeV}) \sim 1 \times 10^{-11}$             & \cite{Ogio1993}  \\ 
      1993              & CANGAROO-I    & \psr        & n/a       & $F(> 3.2 \pm 1.6\,\mathrm{TeV}) < 8.0 \times 10^{-13}$    & \cite{Yoshikoshi2009}  \\
      1993--1994        & CANGAROO-I    & \psr        & n/a       & $F(> 3.2 \pm 1.6\,\mathrm{TeV}) < 6.1 \times 10^{-13}$    & \cite{Yoshikoshi2009}  \\
      1995              & CANGAROO-I    & \psr          & n/a       & $F(> 3.2 \pm 1.6\,\mathrm{TeV}) < 8.9 \times 10^{-13}$    & \cite{Yoshikoshi2009}  \\
      1997              & CANGAROO-I    & \psr          & n/a       & $F(> 1.8 \pm 0.9\,\mathrm{TeV}) < 4.1 \times 10^{-12}$    & \cite{Yoshikoshi2009}  \\
      1998              & CANGAROO-I    & \psr          & n/a       & $F(> 2.7 \pm 1.4\,\mathrm{TeV}) < 1.3 \times 10^{-12}$    & \cite{Yoshikoshi2009}  \\
      1993--1994        & BIGRAT        & \psr          & n/a       & $F(> 0.5\,\mathrm{TeV}) < (7.0 \pm 0.7) \times 10^{-11}$  & \cite{Rowell1998} \\ 
      1996              & Durham Mark 6 & \psr          & n/a       & $F(> 0.3\,\mathrm{TeV}) = (3.9 \pm 0.7) \times 10^{-11}$  & \cite{Chadwick_1998} \\
      2000--2001        & CANGAROO-II   & \psr          & n/a       & n/a                                                     & \cite{Kushida2003} \\
      2003              & H.E.S.S.      & \snr\ center  & 0.6\degr  & $F(> 0.50\,\mathrm{TeV}) < 7.6 \times 10^{-12}$           & Appendix A \\ 
      2003              & H.E.S.S.      & \snr\ center  & 0.6\degr  & $F(> 0.60\,\mathrm{TeV}) < 6.3 \times 10^{-12}$           & Appendix A  \\
      2004--2007        & CANGAROO-III  & \psr          & 0.26\degr & $F(> 1\,\mathrm{TeV}) = (3.0 \pm 0.6) \times 10^{-12}$     & \cite{Enomoto2009} \\
      2004--2007        & CANGAROO-III  & \psr          & 1.0\degr  & $F(> 1\,\mathrm{TeV}) \approx 2.2 \times 10^{-11}$         & \cite{Enomoto2009} \\
      2007              & H.E.S.S.      & \psr          & 0.1\degr  & $F(> 0.6\,\mathrm{TeV}) < 3.3 \times 10^{-13}$             & Sect.\ 3 \\ 
      2007              & H.E.S.S.      & \snr\ center  & 0.6\degr  & $F(> 0.6\,\mathrm{TeV}) \approx 6.5 \times 10^{-12}$       & Sect.\ 3 \\ 
      \hline
    \end{tabular}
    \caption{Summary of the efforts to observe \psr\ in the VHE $\gamma$-ray domain. The CANGAROO upper limits (ULs) are at a 95\% 
      confidence level (CL), the BIGRAT UL is at 3~$\sigma$ (99.7\% CL), and the H.E.S.S.\ ULs are at a 99\% CL.
      The CANGAROO-I integral flux based on the 1992 dataset \citep{Ogio1993} likely 
      suffered from systematics similar to those that affected the 1993 data, which has since been re-analysed along with an analysis of the previously unreleased 
      1994--1998 CANGAROO-I data; the ULs assume a Crab-like spectral index $\Gamma = -2.5$ \citep{Yoshikoshi2009}.  Only the latest, revised results are shown in this table;
      see Sect.~1 for further discussion.
      The BIGRAT UL is subject to an additional $\pm$~50\% systematic uncertainty \citep{Rowell1998}.
      The integral flux from the 2000--2001 CANGAROO-II data analysis was not disclosed but was claimed to confirm previous results \citep{Kushida2003}.
      The 2003 H.E.S.S.\ ULs are based on the re-analysis presented in Appendix~A; the first UL (row 10) assumes $\Gamma = -2.5$ for comparison to the CANGAROO ULs,
      while the second UL (row 11) assumes $\Gamma = -2.0$ for comparison to the 2007 H.E.S.S.\ detection.
      The CANGAROO-III fluxes are from the ON-OFF analysis presented in \cite{Enomoto2009}.
      The 2007 H.E.S.S.\ results are described in Sect.~3, where the center of \snr\ is also defined; the point-source UL assumes $\Gamma = -2.5$.
      }
    \label{TableHistory}
  \end{center}
\end{table*}

In the VHE domain, both the pulsar
and the SNR have been observed using ground-based, imaging atmospheric-Cherenkov
telescopes (IACTs). The findings of
the various observations are, however, not fully consistent (see Table\,\ref{TableHistory}). The 
CANGAROO (Collaboration of Australia and Nippon (Japan) for a Gamma Ray Observatory in the Outback) Collaboration
reported the detection of steady emission, coincident with the pulsar, using the 3.8 m CANGAROO-I telescope in 1992--1993
\citep{Kifune_1995}.
They measured an integral flux $F$($\gtrsim$~1~TeV)~$\approx$~$1 \times 10^{-11}$~ph~cm$^{-2}$~s$^{-1}$,
equivalent to $\sim$44\% of the Crab Nebula 
flux\footnote{The Crab Nebula spectrum published in \cite{Aharonian_2006_Crab} is adopted as the reference Crab spectrum 
throughout this paper.}. 
However, the CANGAROO Collaboration recently undertook a comprehensive re-analysis of their archival CANGAROO-I data
and no longer find a signal;
instead, they calculate an upper limit (UL; here, at 95\% confidence level and assuming a spectral index of $-$2.5)
to the integral flux of $F$($\gtrsim$~3.2~TeV)~$<$~$8.0 \times 10^{-13}$~ph~cm$^{-2}$~s$^{-1}$ ($\sim$24\% Crab)
\citep{Yoshikoshi2009}.
The 4-m BIGRAT (BIcentinnial Gamma RAy Telescope) telescope
\citep{Rowell1998} also observed the pulsar in 1993--1994 and reported a compatible UL. Observations in 1996
with the Durham Mark 6 telescope \citep{Chadwick_1998} appeared to confirm the earlier CANGAROO-I detection,
with a reported integral flux
that was compatible within the large systematic uncertainties ($\pm$30\% for CANGAROO-I and $\pm$50\% for the Mark 6).
Further observations with the CANGAROO-II telescopes in 2000--2001 again seemed to validate the detection
\citep{Kushida2003}.  However, when
the H.E.S.S. (High Energy Stereoscopic System) Collaboration observed the pulsar in 2003 during its commissioning phase,
they did not detect any significant VHE
$\gamma$-ray emission from \psr\ or its vicinity.  The derived UL (99\% confidence level)
on the integral flux from an extended 
region encompassing the SNR was found to be $F$($>$~0.5~TeV)~$<$~$3.5 \times 10^{-12}$~ph~cm$^{-2}$~s$^{-1}$ 
($\sim$5\% Crab) \citep{Aharonian_2005_1706}, in stark disagreement with all of the previous findings (see also the 
Appendix).  
Shortly thereafter, preliminary analysis of stereo observations with the 4~$\times$~10-m
CANGAROO-III telescope array also disagreed with
the initial CANGAROO-I detection and resulted in an UL at the pulsar position (95\% confidence level) of 
$F$($\gtrsim$~0.6~TeV)~$\lesssim$~$5 \times 10^{-12}$~ph~cm$^{-2}$~s$^{-1}$ ($\sim$10\% Crab)
\citep{Tanimori2005}, which agreed with the H.E.S.S.\ results at the time.  

In 2007, H.E.S.S.\ followed up on their initial result by carrying out additional dedicated observations of the pulsar,
now utilizing the superior sensitivity of the fully-operational H.E.S.S. telescope array.  
This campaign resulted in the discovery of extended emission from the vicinity of \psr\ and \snr, 
with preliminary results published in \cite{Hoppe2009}.
The latest results from CANGAROO-III also indicate the presence of an extended source of
VHE $\gamma$-ray emission from the vicinity of the pulsar, although the spectrum and morphology of the emission
vary considerably depending on the method used for background subtraction \citep{Enomoto2009}.  For example,
integrating within 1.0\degr\ of the pulsar position and using an ON-OFF background method (see Sect.~2.3),
they find a Crab Nebula-level integral flux.
In this paper, we present new VHE data on \psr\ and \snr\ which was obtained
during H.E.S.S.'s 2007 observational campaign.

\section{H.E.S.S. Observations and Analysis}

\begin{figure*}[t!]
\begin{minipage}[t]{0.5\textwidth}
   \includegraphics[scale=0.48]{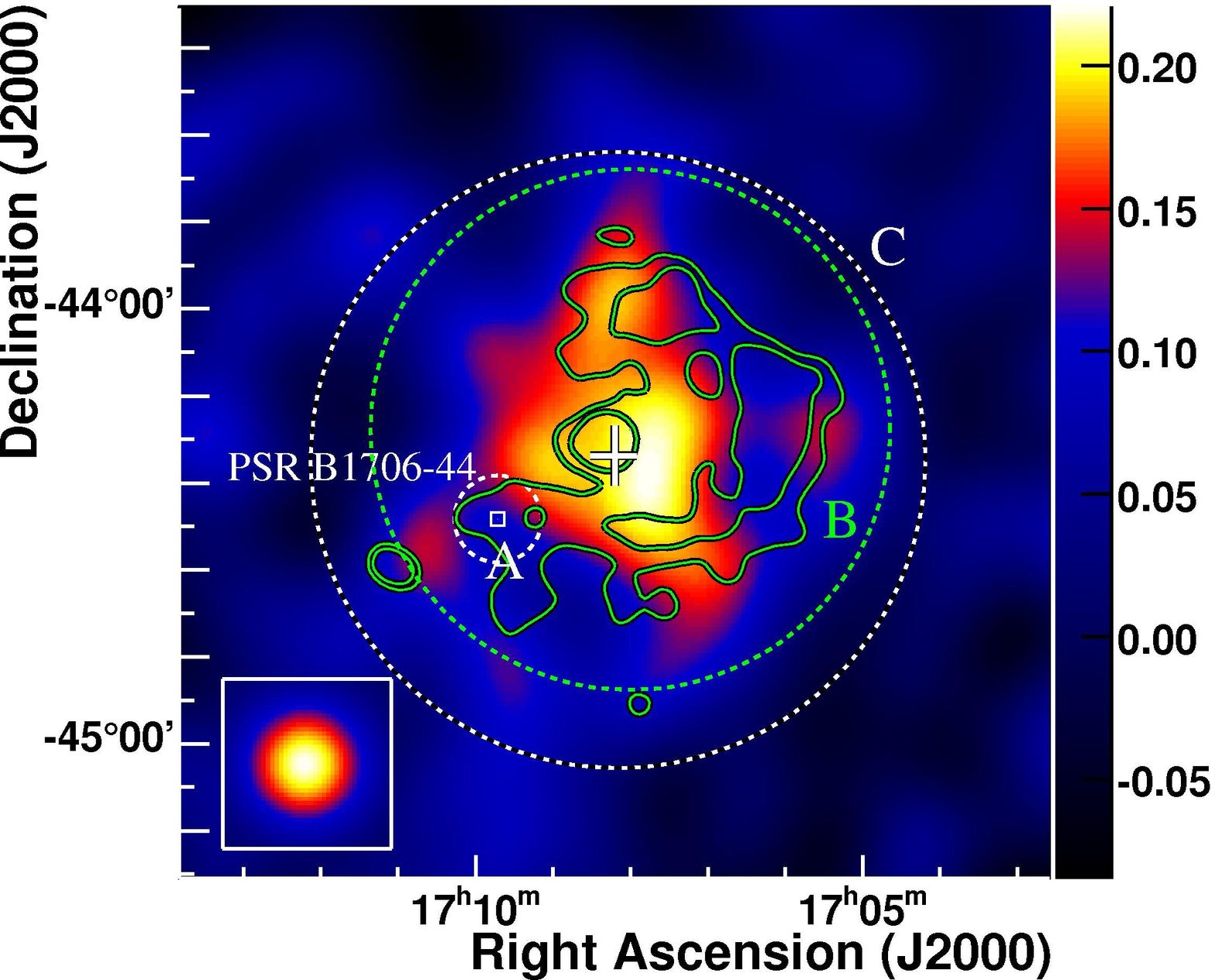}
 \end{minipage}
\hspace{0.6cm}
 \begin{minipage}[t]{0.5\textwidth}
   \includegraphics[scale=0.48]{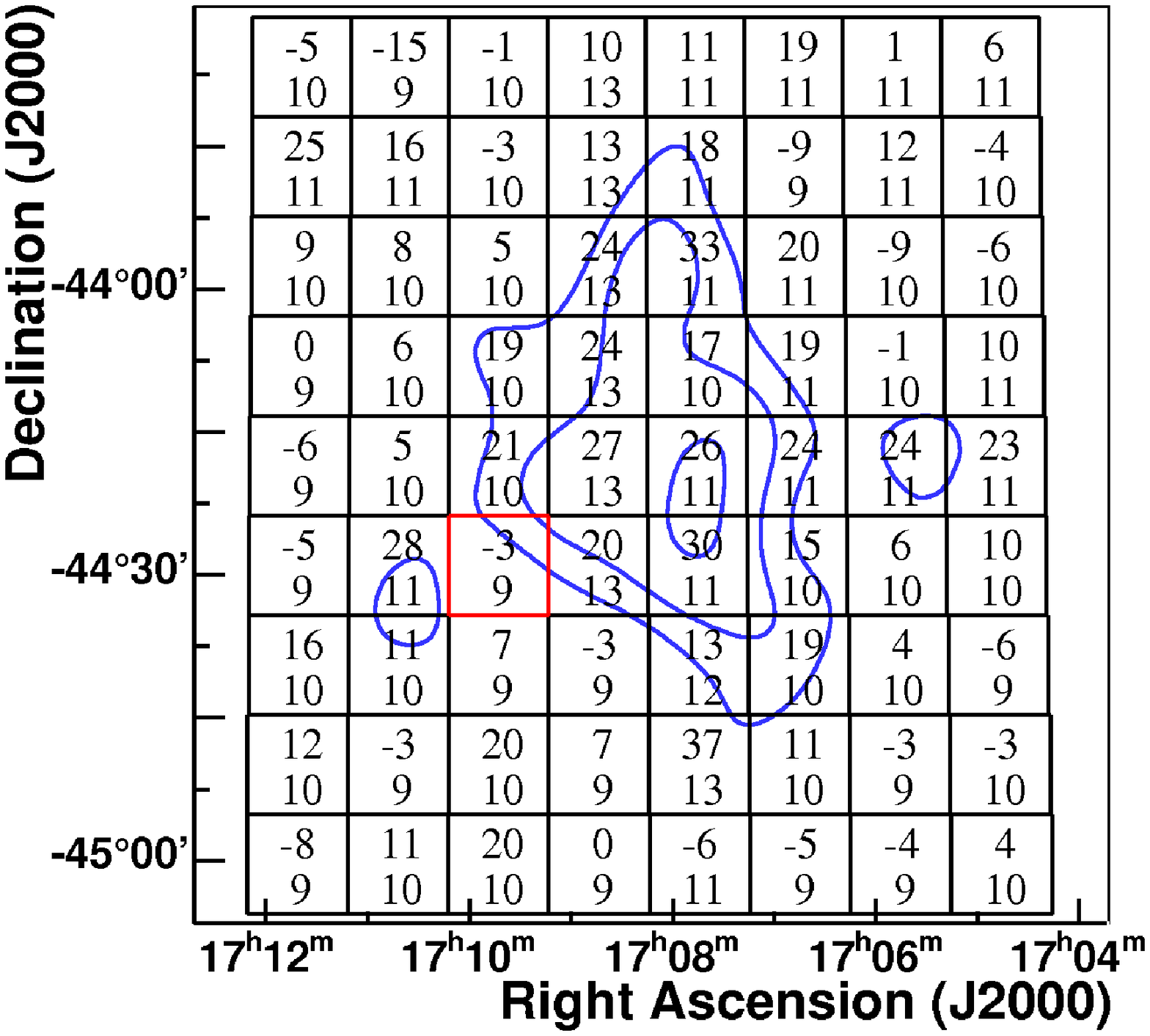}
 \end{minipage}
 \caption{{\it Left:} Image of the VHE $\gamma$-ray excess (in units of $\gamma$-rays~arcmin$^{-2}$) from
   \hessj, smoothed with a 2D Gaussian with
   a width $\sigma$~$=$~0.10\degr. The blue-to-red color transition is chosen to reduce the appearance
   of features which are not statistically significant. The white cross is located at the 
   best-fit position of the center-of-gravity of the emission and its size
   represents the statistical error of the fit. The small and large dotted white circles,
   labeled A and C, respectively, denote the regions used for
   spectral analysis. The \emph{a priori} defined Region B, from which the detection significance
   was calculated, is represented by a dotted green circle. The three regions are
   summarized in Table~\ref{RegionStatsTable}. The position of the
   pulsar \psr, at the center of region A, is marked by a square. The inset 
   (bottom-left corner) shows the point-spread function of the
   H.E.S.S. telescope array for this particular dataset, smoothed in the same manner as
   the excess image. Radio contours of constant intensity, as seen at 330~MHz
   with the Very Large Array (VLA), are shown in green. The radio data were
   smoothed with a Gaussian of width $\sigma$~$=$~0.03\degr. 
   The local maximum in the radio contours at the center of the image is largely due to PMN\,J1708$-$4419,
   an extragalactic object seen in projection (see Sect.\ 4.3).
   {\it Right:} Gamma-ray excess in quadratic bins of 0.175\degr\ width.
   The upper number in each bin is the excess
   summed within this bin, and the lower number is the corresponding statistical error.
   The blue contours correspond to a smoothed
   excess of 0.14, 0.17, and 0.21 $\gamma$-rays~arcmin$^{-2}$, taken from
   the image on the left. The red-rimmed bin is centered on the
   pulsar position. Note the different field-of-view used in the two figures.}
\label{fig1}
\end{figure*}

\subsection{The H.E.S.S. Telescope Array}
H.E.S.S. is an array of four
IACTs, dedicated to the observation
of VHE $\gamma$-rays. The array has been operating since December 2003 in the
Khomas Highlands of Namibia; its location in the southern hemisphere
(23\degr16'17''~S) allows observations of the inner Galaxy at reasonably low
zenith angles.
Each telescope is equipped with a
tessellated, spherical mirror with a total area of 107~m$^{2}$ and a camera
comprised of 960 photomultiplier tubes, covering a field-of-view (FoV) 
5\degr\ in diameter. The telescopes are situated on a square with
sides of 120~m length and operated in \emph{stereo trigger} mode \citep{Funk2004}, which requires
at least two telescopes to trigger the detection of an extended air shower (EAS).
This stereoscopic approach results in an angular
resolution $\lesssim 0.1\degr$ per event, an energy resolution
of $\sim$15\% (on average), and an improved background rejection
\citep{Aharonian_2006_Crab}. The H.E.S.S. array can detect point
sources near zenith at flux levels of $\sim$1\% of the Crab Nebula flux
with a statistical significance of 5~$\sigma$ in 25~h of
observations, or less if advanced techniques are used for EAS image
analysis \citep{NGC253}.
Its large FoV and off-axis sensitivity not only make it
ideally suited for surveying the Galactic Plane \citep{Aharonian2005GPS,Aharonian_2006_SurveyII,Chaves2008GPS}, but
also for studying extended sources like \hessj.

\subsection{VHE $\gamma$-ray Observations}
The region of interest, which includes \psr\ and SNR
\snr, was observed with the full four-telescope H.E.S.S. array in 
2007. The observations were dedicated to search for VHE $\gamma$-ray
emission from the pulsar and were therefore taken in \emph{wobble} mode
centered on its position in the radio
($\alpha_{2000}$~$=$~17$^{\mathrm{h}}$09$^{\mathrm{m}}$42.73$^{\mathrm{s}}$,
$\delta_{2000}$~$=$~$-$44\degr29\arcmin08.2\arcsec; \cite{Wang_2000}). In
this observation mode, the array is pointed toward a position offset
from the source of interest to allow simultaneous background
estimation. Observations of 28-min duration were taken, alternating
between offsets of $\pm$0.7\degr\ in declination and right
ascension. After standard quality selection \citep{Aharonian_2006_Crab} to remove data
affected by unstable weather conditions or hardware-related problems, the total
live-time of the dataset is $\sim$15~h. The zenith angle of the
observations ranges from 20\degr\ to 30\degr, with a mean
of 24\degr.  We only use data from the 2007 observations of \psr, 
because at that time the full four-telescope array was in operation along with the central stereo trigger system, 
resulting in a higher sensitivity compared to earlier observations in 2003 \citep{Aharonian_2005_1706} 
when H.E.S.S. was in its commissioning phase, with only two telescopes and no central trigger (see also the Appendix).

\subsection{Analysis Methods} 
The dataset was analyzed using the Hillas second moment method \citep{Hillas1985}
and the H.E.S.S. standard analysis described in \cite{Aharonian_2006_Crab}.
For $\gamma$-hadron separation,
{\it hard cuts} were used, which require a minimum of
200 photoelectrons (p.e.) to be recorded per EAS image. Compared to
{\it standard cuts} (80 p.e.), this relatively strict requirement results in
better background rejection and an improved angular resolution but
also in an increased energy threshold (560~GeV for this
dataset). The time-dependent optical response of the
system was estimated from the Cherenkov light of single muons passing
close to the telescopes \citep{Bolz_2004}.
Three different background estimation procedures
\citep{Berge_2007} were used in this analysis. 

For 2D image
generation, the \emph{ring background method} \citep{Berge_2007} was used with a mean ring radius
of 0.85\degr. Since this method includes an energy-averaged model for the
camera acceptance to account for the different offsets of the signal and
background regions from the camera center, it was not used for
spectral extraction. The \emph{reflected region background method} \citep{Berge_2007} was instead used
to measure the flux from the pulsar position. 

Since the observations of PSR\,B1706$-$44 were performed in \emph{wobble} mode (see Sect.~2.2),
half are actually pointed inside the extended emission from HESS\,J1708$-$443, which was 
not known to exist at that time.  Therefore, for spectral
extraction from extended regions which also enclose the pointing
positions of the telescopes, the background was estimated 
using the \emph{ON-OFF background method} \citep{Berge_2007}, where
off-source (OFF) data taken is taken from extragalactic regions of the sky where no $\gamma$-ray
sources are known.
To match the observing conditions between on-source (ON)
and OFF data, the two observations had to be taken within
six months of each other and at similar zenith angles. 
The \emph{ON-OFF background method} was also used for the analysis of \object{Vela Junior}
\citep{Aharonian2005_VelaJr}.
The normalization
between ON and OFF observations \citep{Berge_2007} was calculated from the total
event number in the two observations, excluding regions with significant VHE
$\gamma$-ray signal.  The background is thus normalized in an approximately ring-shaped 
region (depicted in Fig.~3) with inner radius 1.0\degr\ and outer radius 2.5\degr, excluding a small region
which overlaps the known source \object{HESS\,J1702$-$420} \citep{Aharonian_2008_Dark}.
With this background normalisation, the analysis is obviously only sensitive to a localized excess
of $\gamma$-rays but not to emission which would be more or less uniform across the entire H.E.S.S.\ FoV.

\section{Results}

\begin{table*}[]
  \begin{center}
    \begin{tabular}{c c c c c c c c c c}
      \hline\hline
      Region&\multicolumn{2}{c}{Center}&Radius&N$_{\mathrm{on}}$&N$_{\mathrm{off}}$&$\alpha$&Excess&Significance&Integral flux ($>$ 0.6 TeV)\\
      &$\alpha_{2000}$ &$\delta_{2000}$ & [\degr] & & & & & [$\sigma$] & [ph cm$^{-2}$ s$^{-1}$] \\ 
      \hline
      A &17$^{\mathrm{h}}$09$^{\mathrm{m}}$42.73$^{\mathrm{s}}$ &$-$44\degr29\arcmin8.2\arcsec
      &0.10 &71 &717 &0.11 &$-$9.4$^{+9.2}_{-8.7}$ &
      $-$1.0 & $< 3.3 \times 10^{-13}$ \\ [0.5ex]
      B &17$^{\mathrm{h}}$08$^{\mathrm{m}}$ &$-$44\degr16\arcmin48\arcsec &0.60 &3180
      &2488 &1.06 &543$^{+77}_{-77}$ & 7.0 & $= 6.5 \times 10^{-12}$ \\ [0.5ex]
      C &17$^{\mathrm{h}}$08$^{\mathrm{m}}$11$^{\mathrm{s}}$ &$-$44\degr20\arcmin&0.71 &4243 &3425 &1.06 &615$^{+90}_{-90}$ 
        & 6.8 & $= 6.9 \times 10^{-12}$ \\ \hline
     \end{tabular}
     \caption{Event statistics for Regions A, B, and C. The center and
       the radius of each circular on-source (ON) region is given in columns 2--4. For Region A,
       the background was extracted from off-source (OFF) regions in the same field-of-view, while
       for Regions B and C, it was estimated from observations of separate OFF regions.
       Due to the smaller extent of Region A, more OFF
       regions could be used, which resulted in a smaller normalization
       factor $\alpha$ than for Regions B and C. The number of events in the ON and OFF integration 
       regions, N$_{\mathrm{on}}$ and
       N$_{\mathrm{off}}$, respectively, are given in columns 5 and 6.
       The significance (column 7) was calculated following the approach
       of \cite{LiMa_1983}. 
       The integral flux (or UL thereof) for each region is given in column 8.
       Note that the statistics presented here
       were obtained from a dataset comprised only of observations in 2007, which does not
       overlap with the one used in \cite{Aharonian_2005_1706}.}
    \label{RegionStatsTable}
  \end{center}
\end{table*}

\begin{figure}[t!]
 \begin{center}
    \includegraphics[width=0.49\textwidth]{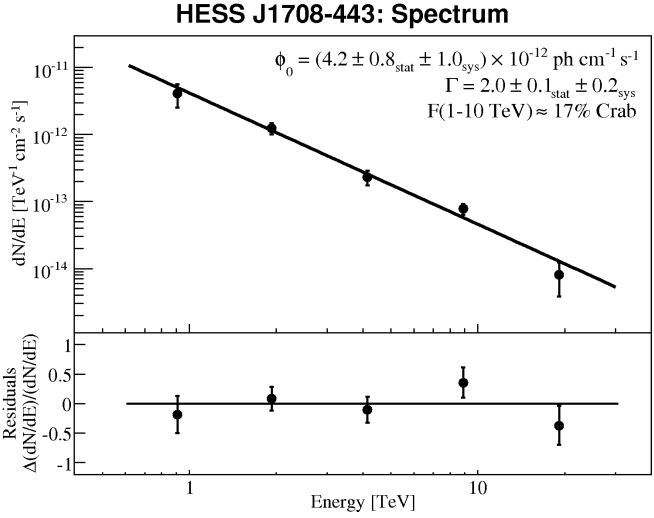}
    \caption[Differential energy spectrum of
    \hessj]{Differential energy spectrum of
      \hessj, extracted from Region C (see Table~\ref{RegionStatsTable}).
      The solid line shows the result of a power-law fit. The error bars denote 1-$\sigma$
      statistical errors.  
      The bottom panel shows the residuals of the power-law fit.
      Events with energies between 0.6 and 28~TeV
      were used in the determination of the spectrum, and the minimum significance per bin is 1 $\sigma$.}
 \label{figSpectrum}
\end{center}
\end{figure}

Two different circular regions were defined \emph{a priori} in order to reduce the number of trials
during a search for statistically-significant VHE $\gamma$-ray emission. Since other IACTs
have reported point-like emission from the pulsar
position, one of these regions, hereafter Region A, is centered at this position and has a radius of
0.10\degr, which is the standard radius used to search for point sources in the
H.E.S.S.\ standard analysis. The second region, hereafter Region B, is
identical to the region referred to as the \emph{Radio arc} in \cite{Aharonian_2005_1706};
it is centered at the approximate apparent center of the radio arc
($\alpha_{2000}$~$=$~17$^{\mathrm{h}}$08$^{\mathrm{m}}$,
$\delta_{2000}$~$=$~$-$44\degr16\arcmin48\arcsec; as defined in \cite{Aharonian_2005_1706}) and has a radius of
0.60\degr\ in order to enclose the entire radio structure.  

No statistically-significant emission is observed
from the pulsar position (Region A); therefore, an upper limit of 14.8 excess
$\gamma$-ray events is calculated at a 99\% confidence level,
following the unified approach of \cite{Feldman_1998}. From Region~B,
however, a clear signal is detected with 543 excess $\gamma$-rays and a 
significance of 7.0 $\sigma$. The measured
signal corresponds to a flux $\sim$13\% that of the Crab Nebula
above 0.6~TeV. Table~\ref{RegionStatsTable} summarizes the event
statistics for Regions A and B.

Figure~\ref{fig1} (left) presents an image of the VHE $\gamma$-ray
excess in the 2\degr~$\times$~2\degr\ region around the source, smoothed with a Gaussian of width
0.09\degr\ to reduce statistical fluctuations. 
The smoothing
radius is chosen to be on the same scale as the H.E.S.S. point-spread function 
(PSF; 68\% containment radius $\sim 0.1$\degr), 
so that resolvable morphological features are largely maintained.
The emission clearly
extends beyond the PSF, which is depicted in the lower left
inlay of Fig.~\ref{fig1} (left).  Figure~\ref{fig1} (right) shows the number of excess events
within the emission region along with their statistical errors, in
quadratic bins of 0.175\degr\ width, without smoothing.
This figure demonstrates that the current statistics do not permit a detailed study of the source morphology.
However, the lack of a significant VHE $\gamma$-ray excess at the position of the pulsar is clear in both
figures.

The centroid of the new H.E.S.S. source is determined by fitting the unsmoothed $\gamma$-ray excess 
image with a radially-symmetric
Gaussian profile ($\phi$~$=$~$\phi_0$~$e^{-r^{2}/(2\sigma^{2})}$)
convolved with the H.E.S.S.\ PSF (0.07\degr\ for this analysis).  The centroid of the best fit is at  
$\alpha_{\mathrm{J2000}} = 17^{\mathrm{h}}08^{\mathrm{m}}11^{\mathrm{s}} \pm 17^{\mathrm{s}}$ and
$\delta_{\mathrm{J2000}} = -44\degr20\arcmin \pm 4\arcmin$ 
($\ell \sim 343.06\degr, b \sim -2.38\degr$).
The pointing precision of the H.E.S.S.\ telescope
array is 20\arcsec\ \citep{Gillessen2005}, which adds an additional systematic uncertainty. 
The combined errors are reflected in the size of the cross
in Fig.~\ref{fig1} (left). Consequently, the new VHE $\gamma$-ray source is
designated \hessj. The fit also gives the source's intrinsic Gaussian
width $\sigma = 0.29\degr \pm 0.04\degr_{\mathrm{stat}}$.

Spectral analyses were performed for two regions, Region A, which was 
introduced above, and Region C, which is centered on the centroid (i.e. best-fit position)
and has a radius of 0.71\degr\ (see Table~\ref{RegionStatsTable}).
The size of Region C represents an
$\sim$95\% enclosure of the excess, chosen as a compromise between
an optimal signal-to-noise ratio and independence of source
morphology. Both regions are indicated by dashed circles in
Fig.~\ref{fig1} (left). From Region A, an integral flux limit of
$F$($>$~0.6~TeV)~$<$~3.3~$\times$~10$^{-13}$~ph~cm$^{-2}$~s$^{-1}$ 
was derived with a 99\% confidence level according to \cite{Feldman_1998}, assuming
that the underlying $\gamma$-ray spectrum follows a power law with photon
index $\Gamma$~$=$~2.5, an index close to that of the Crab Nebula
\citep{Aharonian_2006_Crab}.  This upper limit corresponds to $\sim$1\% of the
flux of the Crab Nebula in the same energy range.

The energy spectrum of the entire source is extracted from Region
C. Within the large integration circle, 615~excess $\gamma$-ray 
events were found,
corresponding to a statistical significance of 6.8~$\sigma$
(pre-trials). The differential spectrum (Fig.~\ref{figSpectrum}) is well-described by a power law
$\phi$~$=$~$\phi_0$~($E$~$/$~1~TeV)$^{-\Gamma}$ with a spectral photon index
$\Gamma$~$=$~2.0~$\pm$~0.1$_{\mathrm{stat}}$~$\pm$~$0.2_{\mathrm{sys}}$
and a flux normalization at 1 TeV of
$\phi_0 = (4.2 \pm 0.8_{\mathrm{stat}} \pm 1.0_{\mathrm{sys}}) \times 10^{-12}$~cm$^{-2}$~s$^{-1}$~TeV$^{-1}$.
The integral flux $F$(1--10~TeV)~$=$~$3.8$~$\times$~10$^{-12}$~ph~cm$^{-2}$~s$^{-1}$ is $\sim$17\% of the Crab Nebula
flux in the same energy range. The extracted flux points from the
extended emission and the fitted power law are shown in
Fig.~\ref{figSpectrum}.
The results presented above have been cross-checked, using an independent calibration of the raw data and an 
alternative analysis chain.
The cross-checks included a spectral analysis using the \emph{reflected region background method} \citep{Berge_2007},
which requires observations to be centered outside of the emissive region and thus used only half of the available
dataset.  All cross-checks confirmed the primary results within the stated statistical uncertainties.

The most recent observations and analysis by CANGAROO-III also give an indication of extended emission in the vicinity of
\psr\ \citep{Enomoto2009}.  However, their results
differ significantly from those given in this paper.  
For example, the morphology of the VHE $\gamma$-ray excess reported by \cite{Enomoto2009},
using an ON-OFF background technique,
is that of a source centered roughly at the pulsar position, as opposed to \hessj., whose centroid is clearly
offset from the pulsar.
Furthermore, CANGAROO-III measures a Crab Nebula-level integral flux (above 1~TeV) within 1.0\degr\ of the pulsar,
which is inconsistent with the $\sim$18\% Crab flux measured by H.E.S.S. in the same energy range.
The difference is possibly due to the exact methods used for background
subtraction; in the H.E.S.S.\ analysis, the OFF data are normalized to
source-free regions of the ON data, because the background can vary
significantly depending on the observing conditions.

\section{Origin of the VHE $\gamma$-ray Emission}

While a superposition of a relic PWN created by \psr\
and SNR \snr\ cannot be excluded, each of these objects individually could
account for the observed VHE $\gamma$-ray emission. The possible
associations with \hessj. will be discussed in the following sections and
both leptonic and hadronic scenarios will be considered.

\subsection{A Relic Nebula from \psr}
The pulsar \psr, which has a high spin-down luminosity 
$\dot{E}$~$=$~3.4~$\times$~10$^{36}$~erg~s$^{-1}$, is energetic enough to power the
observed VHE $\gamma$-ray emission, 
which has a luminosity between 1 and 10~TeV of
$L_{\gamma} \approx 9.9 \times 10^{33} (D / 2.3 \mathrm{kpc})^2$~erg~s$^{-1}$. 
The apparent conversion efficiency from rotational energy to
$\gamma$-rays in this energy range can be defined as
$\epsilon_{\mathrm{1-10~TeV}}$~$\equiv$~$L_{\gamma}$~$/$~$\dot{E}$ and for this case is 
$\sim$0.3\%, 
compatible with the efficiencies ($\lesssim$~10\%) of other VHE $\gamma$-ray sources which have
well-established associations with PWN \citep{Gallant2007}.
The
projected size of \hessj\ corresponds to a physical size 
of $\sim$12~$(D / 2.3 \mathrm{kpc})$~pc (68\% containment radius). 
These characteristics suggest a possible association between the VHE $\gamma$-ray
emission and the PWN of \psr, similar to other PWN/VHE
associations, e.g. \object{Vela~X} \citep{Aharonian_2006_VelaX} and
\object{HESS\,J1825$-$137} \citep{Aharonian_2006_1825II}. 

In a leptonic scenario, the VHE $\gamma$-radiation originates from accelerated
electrons which up-scatter ambient photons to VHE $\gamma$-ray energies via
inverse Compton (IC) scattering.  Compared to the size of the PWN in the radio (radius
$\sim$1.5\arcmin) \citep{Giacani_2001} and the ``bubble'' nebula seen in X-rays (radius
$\sim$1.8\arcmin) \citep{Romani_2005}, the VHE $\gamma$-ray PWN (sometimes referred to as 
a TeV PWN) would be a factor of $\sim$10 larger. Similar differences in size have been observed in
other TeV PWN associations, e.g. HESS\,J1825$-$137
\citep{Aharonian_2006_1825II}, and can be explained by the different
energies, and hence cooling times, of the electrons which emit 
the X-rays and VHE $\gamma$-rays.  Assuming the magnetic field is uniform and that the
average wind convection speeds in the $\gamma$-ray and X-ray emitting zones are both constant and similar,
\cite{Aharonian_2005_1825I} estimate the ratio of sizes
\begin{equation}
  \frac{R_{\gamma}}{R_{\mathrm{X}}}=
  4 \left(\frac{B}{10~\mu \mathrm{G}}\right)^{-\frac{1}{2}}
  \left(\frac{E_{\mathrm{keV}}}{E_{\mathrm{TeV}}} \right)^{\frac{1}{2}}~,
\label{EQ::NebulaSizes}
\end{equation}
where $E_{\mathrm{keV}}$ is the mean energy in X-rays (2~keV) and
$E_{\mathrm{TeV}}$ is the mean energy in VHE $\gamma$-rays (0.9~TeV).
However, in contrast to the PWN of
\object{PSR\,J1826$-$1334}, where a magnetic field strength $B$~$=$~10~$\mu$G was
inferred from X-ray observations \citep{Gaensler_2003},
\cite{Romani_2005} estimated a magnetic field $B$ as strong as 140$^{+210}_{-60}$~$\mu$G
within the 110\arcsec~radius X-ray PWN of \psr, assuming the spectral break
between the extrapolation of radio and X-ray spectra is due to radiative cooling of electrons.
In such a high magnetic
field, electrons that emit keV X-rays have comparable energies to those
that emit TeV $\gamma$-rays and therefore have comparable cooling times as well.
Thus, the TeV PWN should be approximately the same size as the X-ray PWN, i.e. 
it should appear point-like considering the $\sim$5\arcmin~H.E.S.S.
PSF.  Furthermore, given that the ratio of X-ray to VHE
$\gamma$-ray energy flux is determined by the
energy density in magnetic fields and IC target photon fields
(only the cosmic microwave background (CMB) is considered here),
\begin{equation}
\frac{F_{\gamma}}{F_{\mathrm{X}}} \approx 0.1 (0.1 B_{-6})^{-2},
\end{equation}
where $B = 10^{-6} B_{-6}$~G \citep{Aharonian_1997},
the observed X-ray flux $F_{\mathrm{X}} = 3.3 \times 10^{-13}$~erg~cm$^{-2}$~s$^{-1}$ at $E_{\mathrm{keV}} = 1.2$~keV \citep{Becker_1995} can be used
to predict the $\gamma$-ray flux $F_{\gamma}$ at $E_{\mathrm{TeV}} = 167 E_{\mathrm{keV}} B_{-6}^{-1} = 1.4$~TeV
\citep{deJager08} assuming the value of $B$ estimated by \cite{Romani_2005}.  This
results in a predicted $F_{\gamma} (1.4~\mathrm{TeV}) = 1.7 \times 10^{-16}$~erg~cm$^{-2}$~s$^{-1}$,
well below the level observable by H.E.S.S.
Conversely, the absence of VHE $\gamma$-rays from the compact nebula (c.f.\ H.E.S.S.\ UL from Region A 
in Sect.~3)
can be used together with $F_{\mathrm{X}}$ to calculate a lower limit on the magnetic field using Eq.~2.
The resulting limit, $B \ga 2.5~\mu$G, is consistent with the magnetic field estimated by \cite{Romani_2005}.

\begin{figure}[t!]
  \resizebox{\hsize}{!}{\includegraphics{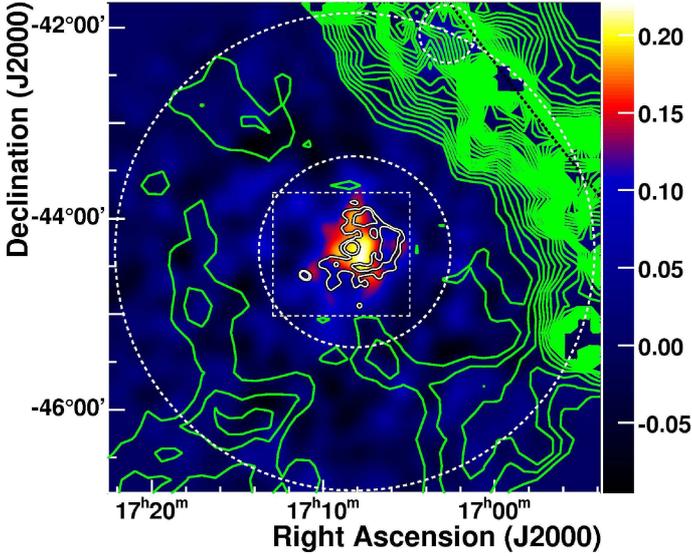}}
  \caption{Large field-of-view (FoV; 5.1\degr~$\times$~5.1\degr) VHE $\gamma$-ray image of
    the region containing \hessj. The Gaussian-smoothed ($\sigma$~$=$~0.10\degr)
    VHE $\gamma$-ray excess from Fig.~\ref{fig1} (left) is
    shown in color. The white contours indicate the intensity
    of the 330~MHz radio emission detected with Very Large Array (VLA) observations (see
    also the green contours in Fig.~\ref{fig1}, left). The dotted white box
    represents the FoV covered by the VLA observations.
    Outside of this region, green contours indicate the lower-resolution 2.4~GHz
    radio continuum data \citep{Duncan_1995} taken with the Parkes
    telescope; these observations have a half-power beamwidth of $\sim$10.4\arcmin.
    This image also shows the approximately ring-shaped region 
      used for normalizing the background in the \emph{ON-OFF background method} (see Sect.~2.3);
      this region is delimited by the large dashed white circles, excluding the known
      TeV source HESS\,J1702$-$420 located toward the Northwest.
    The Galactic plane is also located toward the NW
    and is indicated by a thick black dotted line.}
  \label{fig2}
\end{figure}

One way to reconcile the difference in emission region size
and the high flux of the VHE $\gamma$-ray emission is to assume that the size
of the X-ray PWN is primarily governed by the extent of the
high $B$-field region and that the magnetic field decreases by a large
factor beyond the X-ray PWN. The electrons can then escape from the
high $B$-field region and, by accumulating over a significant fraction
of the lifetime of the pulsar, form a larger nebula which is visible only in
VHE $\gamma$-rays. The synchrotron cooling time of electrons that
up-scatter CMB photons to energies $E_{\gamma}$ is given by
\begin{equation}
  \tau_{\mathrm{synch}} \approx 40 \:
  \Big(\frac{B}{140~\mu\mathrm{G}}\Big)^{-2}\Big(\frac{\mathrm{E}_{\gamma}}{\mathrm{TeV}}\Big)^{-1/2}
  \mathrm{yr}\,.
  \label{EQ::LifeTime}
\end{equation}
In the 140~$\mu$G field inside the X-ray PWN, the cooling time of
up-scattered electrons producing 1~TeV $\gamma$-rays is
$\sim$40~yr. 
Assuming a dominantly advective, rather than diffusive,
transport process, the average flow speed needed to drive
electrons from the pulsar position to the edge of the X-ray PWN ($r~\approx$~110\arcsec)
within 40~yr is $0.1 (D / 2.3 \mathrm{kpc})$~c.
The implied flow speed is reasonable following the arguments of 
\cite{Kennel_1984}, although their model considers the case of the 
symmetric Crab Nebula, which is admittedly a simplification of the asymmetric
PWN considered here.
If the magnetic field within the X-ray PWN was much higher than
140~$\mu$G in the past, when most of the electrons were
emitted, the restrictions on the flow speed would become more
stringent. However, in the low $B$-field region outside the X-ray PWN, the
synchrotron lifetime increases.  Even for a magnetic field strength of
10~$\mu G$, a value about three times as large as the interstellar
magnetic field, the cooling time of the aforementioned electrons is
about 8\,000 yr, almost half of the characteristic age of the
pulsar (17\,500~yr). 

\begin{figure*}[t!]
\begin{minipage}[t]{0.5\textwidth}
  \includegraphics[scale=0.49]{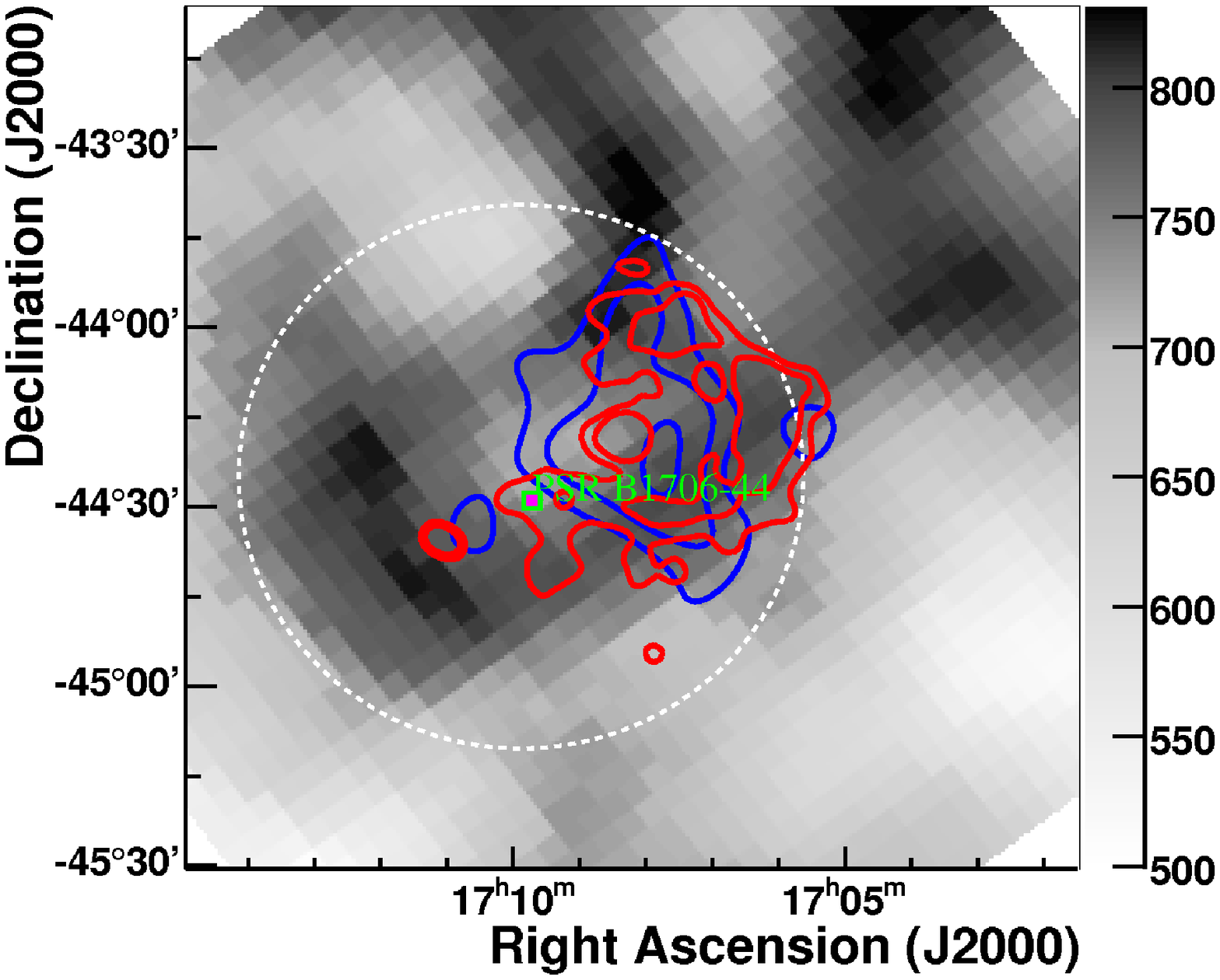}
 \end{minipage}
\hspace{0.1cm}
 \begin{minipage}[t]{0.5\textwidth}
   \includegraphics[scale=0.523]{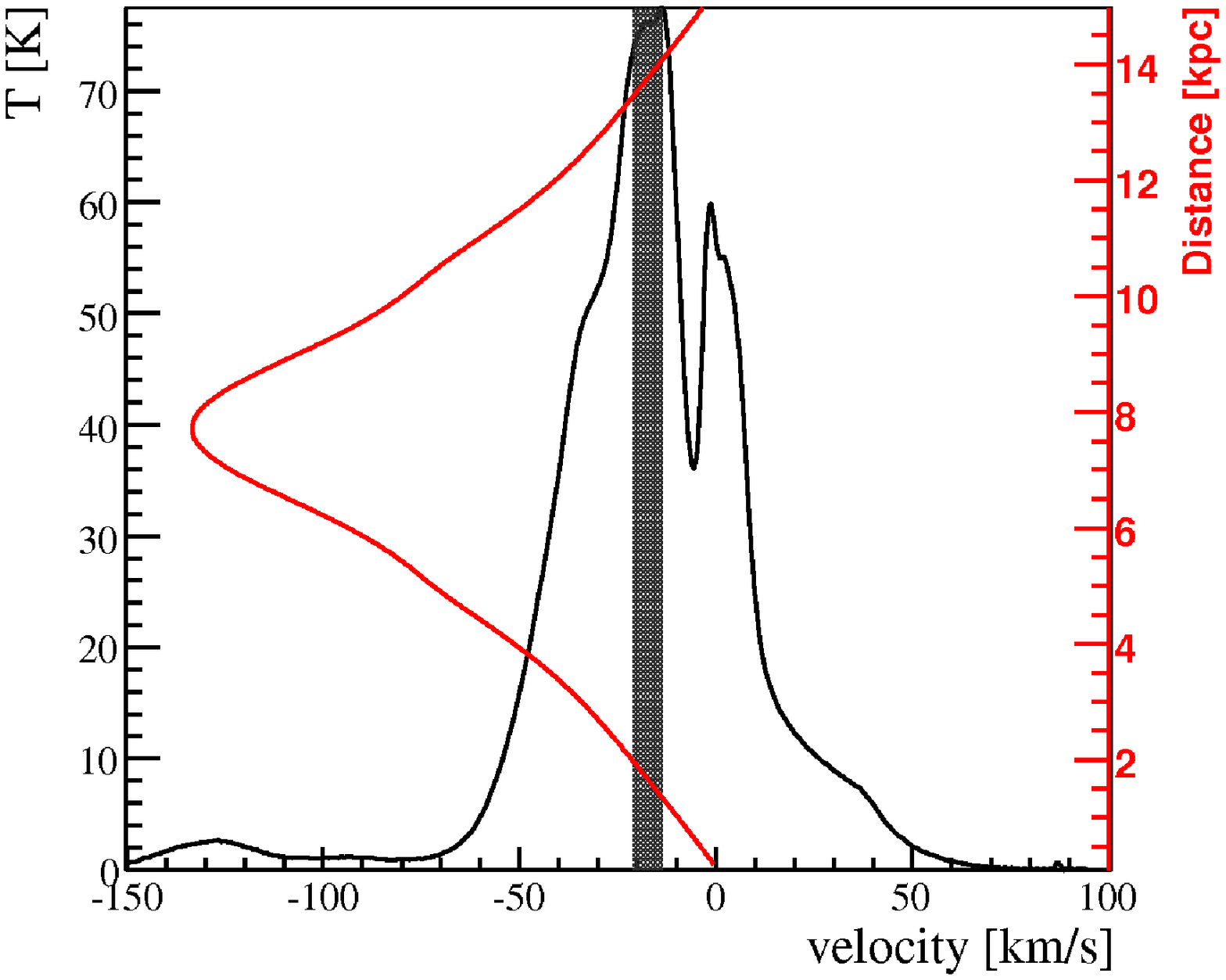}
 \end{minipage}
 \caption{{\it Left:} The grey-scale image shows the intensity of \ion{H}{I} line emission in units of
   K~km~s$^{-1}$, measured by the Parkes radio telescope during the
   Southern Galactic Plane Survey (SGPS) \citep{McClure_2005}. The
   intensities are integrated in the velocity range 
   $-$13.79~km~s$^{-1}$ to $-$21.21~km~s$^{-1}$ (shown as a shaded region in the velocity profile, right),
   corresponding to a
   near/far kinematic distance of 1.6--2.3~kpc / 13.0--13.7~kpc. 
   Contours of the
   Gaussian-smoothed ($\sigma$~$=$~0.10\degr) VHE
   $\gamma$-ray excess are shown in blue. The red contours depict the
   intensity of the radio emission measured by the Very Large Array (VLA)
   at 330~MHz (see also Fig.~\ref{fig1} left). The radio data have been
   smoothed with $\sigma$~$=$~0.03\degr.
   The white circle
   illustrates the integration region for the velocity profile shown
   on the right. {\it Right:} Velocity profile of \ion{H}{I} line emission
   intensity, integrated over the region enclosed by the dashed circle on the
   left. The velocity resolution is 0.08~km~s$^{-1}$. The
   kinematic distance, shown in red, is derived from the velocity using 
   the Galactic rotation curve of \cite{Fich_1989}.}
\label{fig3}
\end{figure*}

This relic TeV PWN scenario does not, however, explain the 
asymmetric morphology of \hessj, in particular its offset
from the pulsar location, nor does it explain the lack of detectable
VHE $\gamma$-ray radiation from the location of the pulsar itself, 
assuming that the pulsar and X-ray PWN are embedded in an extended shell of relic electrons.
Such asymmetries have been observed previously in other TeV PWNs,
e.g. \object{HESS\,J1718$-$385}, \object{HESS\,J1809$-$193}
\citep{Aharonian_2007_twopwn} and HESS\,J1825$-$137
\citep{Aharonian_2005_1825I,Aharonian_2006_1825II}. 
These asymmetries
could be accounted for in two ways: as a direct
result of a high proper motion of the pulsar or as a result of  
a density gradient in the ambient medium.  The density gradient could lead to an asymmetry
in the reverse shock of the supernova, or it could lead to a
different expansion velocity for the TeV $\gamma$-ray emitting 
electrons \citep{Blondin_2001,Swaluw_2001}.  Simulations by \cite{Swaluw_2001}
demonstrate that a displaced PWN can indeed be well-separated from its
pulsar. These explanations are in principle applicable to the case of
\hessj; however, the pulsar's measured scintillation velocity, less than
100~km~s$^{-1}$, renders the first explanation unlikely.  
The latter explanation favors a TeV PWN which is offset toward a low
density region. The available \ion{H}{I} line emission data (see Fig.~\ref{fig3} (left)
and the subsequent section) suggest that this might be the case, although
it is not clear given the complex \ion{H}{I} morphology.

In the preceeding discussion, 
it was assumed that the pulsar dominantly accelerates electrons.  If a considerable fraction of the accelerated
particles are instead hadrons (e.g.\ \cite{Horns2006,Amato2003,Bednarek2003}), the
constraints imposed by the large magnetic field within the X-ray PWN
are removed. In a hadronic scenario, $\pi^0$ mesons are produced by
inelastic interactions between accelerated protons and the ambient gas;
they then decay, emitting VHE $\gamma$-ray photons.  In such a scenario,
the VHE $\gamma$-ray emission would trace the distribution of the target material.  The bright radio
arc, interpreted by \cite{Bock_2002} as the compressed outer
boundary of the former wind-blown bubble, could act as such a target due to its
enhanced density, thereby also explaining the spatial coincidence with the H.E.S.S. source.
Since the proton interaction time is long compared to the age of the pulsar, and assuming that the escape
of protons from the region is sufficiently slow, all protons accelerated since the birth of the pulsar
can contribute to the $\gamma$-ray emission.
However, to account for the
high luminosity of the VHE $\gamma$-ray emission, the pulsar must have a high rotational energy and
must efficiently convert rotational energy into proton acceleration.
The total energy in accelerated protons $W_{\mathrm{p}}$ in the energy
range 10--100~TeV which is necessary to produce the observed $\gamma$-ray
luminosity $L_{\gamma}$ can be estimated from the relation
\begin{equation}
  W_{\mathrm{p}}(10-100~\mathrm{TeV}) \approx \tau_{\gamma} \times L_{\gamma}(1-10~\mathrm{TeV})\,,
\end{equation}
where $\tau_{\gamma} \approx 5 \times 10^{15} (n / \mathrm{cm}^{-3})^{-1}$~s is
the characteristic cooling time of protons through the $\pi^{0}$
production channel. The total energy within the entire proton
population $W_{\mathrm{P}}(\mathrm{tot}) \approx 3 \times 10^{49}~\mathrm{erg}
(n / \mathrm{cm}^{-3})^{-1} (D / 2.3 \mathrm{kpc})^{2}$
is then estimated by
extrapolating the proton spectrum down to 1~GeV assuming the same
spectral shape as the VHE $\gamma$-ray spectrum, i.e.\ a power law with index $\Gamma$~$=$~2.0. 
Assuming that a
fraction $\eta$ of the pulsar's rotational energy $E_{\mathrm{rot}}$ is converted
into the energy within the proton population $W_{\mathrm{P}}(\mathrm{tot})$~$=$~$\eta$~E$_{\mathrm{rot}}$,
then 
\begin{equation}
  \left(\frac{n}{\mathrm{cm^{-3}}}\right) \approx 0.2 \left(\frac{D}{2.3~\mathrm{kpc}}\right)^{2} 
  \eta^{-1} \left(\frac{P_{\mathrm{0}}}{10~\mathrm{ms}}\right)^{2}~,
\end{equation}
where $E_{\mathrm{rot}}$~$=$~$\frac{(2\pi)^{2}}{2}
\frac{I}{P_{\mathrm{0}}^{2}}$ and $I$~$\approx$~1~$\times$~10$^{45}$~g~cm$^2$
is the moment of inertia of the pulsar.
For a distance $D$~$=$~2.3~kpc and an efficiency $\eta$~$=$~0.3, the
initial rotation period $P_0$ has to be as small as 6--12~ms for the ambient
medium density to be in the range $n$~$\approx$~1--5~cm$^{-3}$.  Although pulsars are thought to be 
born with initial periods which are considerably shorter than their present periods, the 
initial rotation period implied for \psr, in the above hadronic TeV PWN scenario, 
is even smaller than that of the Crab pulsar, the only case for which $P_0$ is 
well-determined (19~ms) \citep{Manchester_1977}.

The hadronic PWN scenario is further disfavored by constraints on the proton escape time.
Under the common assumption that the proton 
diffusion coefficient is energy-dependent, i.e.
\begin{equation}
D(E_{\mathrm{p}}) = D_0~(E_{\mathrm{p}}~/~10~\mathrm{GeV})^{\delta}~,
\end{equation}
with a power-law index $\delta \approx 0.5$,
where $D_0$ is the diffusion coefficient at 10~GeV,
one can estimate $D_0$ required to contain protons with energy $E_p = 100$~TeV 
within a certain distance of the pulsar after $t = \tau_{\mathrm{c}}$, 
since the diffusion radius $R_{dif} = 2 \sqrt{D(E) t}$
for timescales less than the proton energy loss time, $t \ll \tau_{\gamma}$.
This containment region can be estimated to have an angular size of $\sim$0.7\degr,
which is the approximate distance between the pulsar and the farthest significant VHE
$\gamma$-ray emission.  At the assumed pulsar distance, this region would have a 
physical size of 28~($D$~$/$~2.3~kpc)~pc.  
The required diffusion coefficient $D_0 \approx 3.4 \times 10^{25}~(D~/~2.3~\mathrm{kpc})^2~\mathrm{cm}^2 \mathrm{s}^{-1}$
is found to be prohibitively low by a factor of 10--100 and can only be reconciled by assuming
a very weak energy dependence, $\delta \lesssim 0.2$.

\onlfig{1}{
\begin{figure*}
\includegraphics[width=0.49\textwidth]{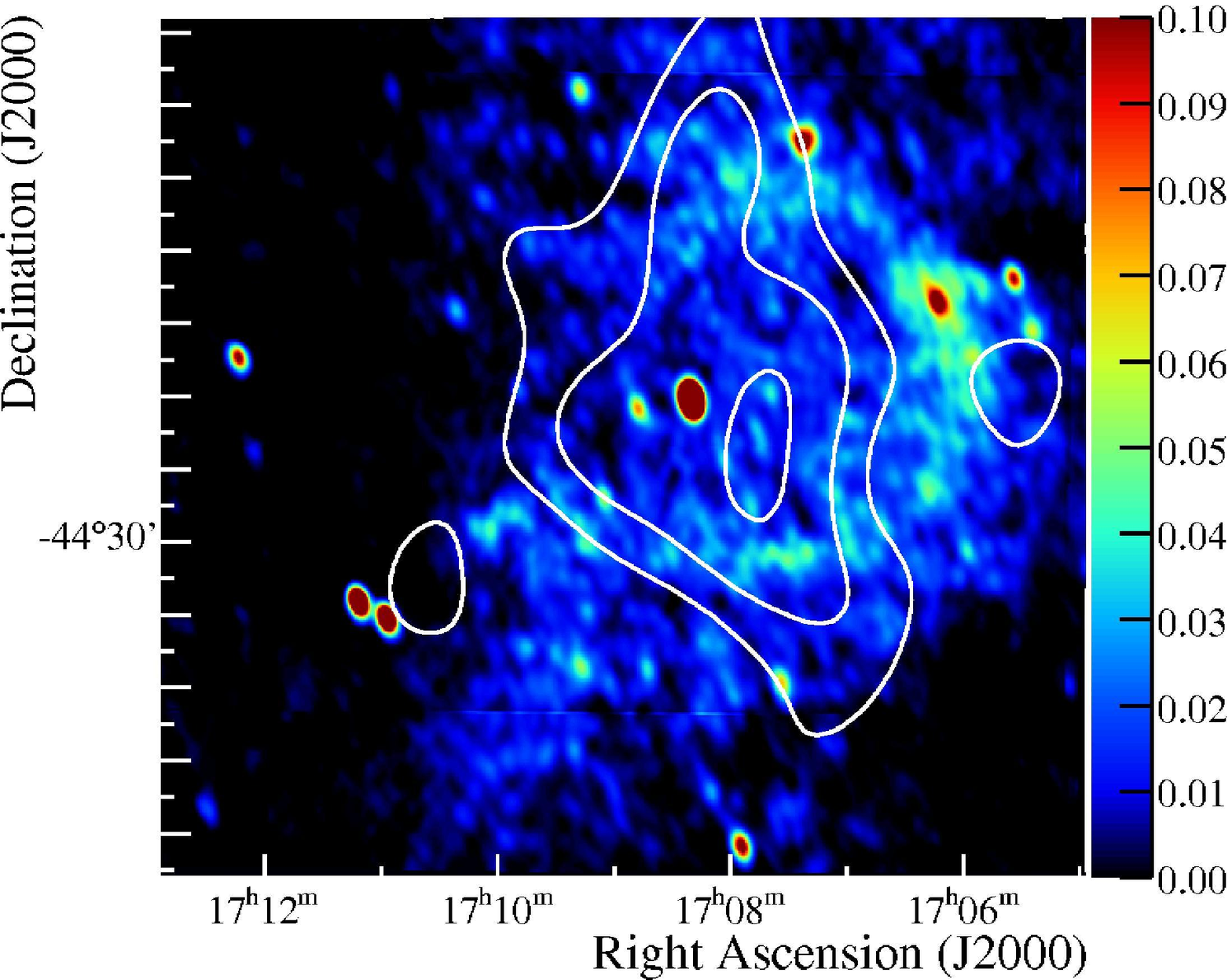}
\caption{Image showing the intensity of the radio emission measured by the Very Large Array (VLA) \citep{Frail_1994}
 at 330~MHz in the vicinity of \psr, smoothed with a Gaussian of width $\sigma$~$=$~0.03\degr. 
 The observations have a half-power beamwidth of $0.03\degr \times 0.015\degr$.
 The radio arc of the partial shell-type SNR \snr\ is clearly visible
 as well as the diffuse emission both inside and outside the arc. 
 The white contours correspond to a smoothed VHE excess of 0.14, 0.17, and 0.21 $\gamma$-rays~arcmin$^{-2}$,
 taken from the image in Fig.~\ref{fig1}, left.
 The horizontal stripes visible at Dec~$= -44\degr40\arcmin$ and Dec~$= -43\degr47.5\arcmin$ are imperfections 
 which resulted from the joining
 of data to form the final wide-field image \citep{Frail_1994}.
 The bright point source at the center of the radio image is PMN\,J1708$-$4419,
 likely an extragalactic object seen in projection (see Sect.\ 4.3).
} \label{figX}
\end{figure*}
}

\subsection{SNR \snr}
The VHE $\gamma$-ray source \hessj\ is
partially coincident with the bright radio arc
and the surrounding diffuse emission of the SNR, visible in the  
330~MHz observations taken with the VLA (see
contours in Fig.~\ref{fig1}, left, and Fig.~\ref{figX}). The centroid of the H.E.S.S.\ source
is consistent with the apparent center of the bright radio arc
($\alpha_{2000}$~$=$~17$^{\mathrm{h}}$08$^{\mathrm{m}}$,
$\delta_{2000}$~$=$~$-$44\degr16\arcmin48\arcsec; as defined in \citet{Aharonian_2005_1706}. 
The extension of the VHE $\gamma$-ray excess (68\% containment radius: 0.29\degr\ $\pm$ 0.04\degr) is compatible with the 
the radius of the radio shell ($\sim$0.27\degr) fit by \citet{Frail_1994} using VLA observations at 90~cm.
The 95\% containment radius (0.71\degr; used for spectral extraction)
of the $\gamma$-ray excess completely encloses the radio shell,
whose approximate boundary was estimated at a radius of $\sim$0.42\degr\ by \citet{Romani_2005} using ATCA 
(Australia Telescope Compact Array) 
observations at 1384~MHz \citep{Dodson_2002}.
Thus, while the majority of the VHE $\gamma$-ray emission is located within the radio shell, 
emission from the shell itself cannot be excluded.
Due to low
statistics in the current VHE dataset, no further conclusions can be made regarding
morphological similarities.
No significant VHE emission was detected
from the spatially-extended, diffuse emission visible farther to the Southeast of
the bright radio arc, seen in the low-resolution 2.4~GHz continuum
radio data \citep{Duncan_1995} shown in Fig.~\ref{fig2}, although the
offset-corrected exposure in this region is very low (between $\sim$4 and 10\,h) since 
all the H.E.S.S.\ observations were centered near the pulsar.
This diffuse radio
emission was interpreted by \cite{Bock_2002} as originating from the
eastern half of the expanding SNR shell, propagating into a
low-density region.

Similar to the potential association with the PWN of \psr,
both leptonic and hadronic scenarios will be considered 
for VHE $\gamma$-ray production. The leptonic scenario suffers from the
non-detection of the SNR at X-ray energies. The VHE $\gamma$-ray
spectrum is hard and extends up to 20~TeV; assuming IC scattering
in the Thomson regime, the electrons which up-scatter CMB
photons to 20~TeV have an energy of $\sim$80~TeV. For a
reasonable magnetic field strength of 5~$\mu$G, these electrons would
emit synchrotron photons with an energy of $\sim$1~keV, i.e.\ photons
within the detectable energy range of current X-ray telescopes.
Unfortunately, this prediction cannot be tested because the X-ray UL calculated by \cite{Becker_1995} using \emph{ROSAT} 
was derived from a relatively small part of the shell; no stringent UL on the X-ray flux
from an extended region within 0.7\degr\ of the
H.E.S.S.\ source can be derived (W.\ Becker, personal communication) due to its large extension 
and the vicinity of the luminous
low-mass X-ray binary (LMXB) \object{4U\,1705$-$440} \citep{Becker_1995}, whose stray light
may be obscuring diffuse X-ray emission from the SNR.  It is also possible that
the X-ray emission is inherently weak and cannot be detected due to the relatively high interstellar absorption
\citep{Becker_1995}.

In the hadronic scenario, synchrotron radiation is expected only
from secondary electrons, and the lack of X-ray detection can easily be
accounted for. Assuming a total energy of 10$^{51}$~erg is released in the
supernova explosion, an acceleration efficiency of $\epsilon$~$=$~0.15 and
a distance $D$~$=$~2.3~kpc, an average proton density of $n$~$\approx$~1.5~cm$^{-3}$ -- a
value slightly larger
than the average Galactic ambient density -- is sufficient
to explain the previously estimated (Sect. 4.1) energy content of
$W_{\mathrm{P}}(\mathrm{tot})$~$\approx$~3~$\times$~10$^{49}$~erg 
within the proton population.

Given the various scenarios that have been proposed to explain the origin of the
bright radio arc, there are many different possibilities as to how the SNR
could be associated with \hessj.  In one scenario, the SNR \snr\
is expanding symmetrically
into the interstellar medium (ISM), and the intensity variations which form the
radio arc are due to local density differences in the ISM.
An association between the SNR and the pulsar \psr,
a controversial scenario which is still debated in the
community \citep[see e.g.][]{Bock_2002,Romani_2005}, would make the
SNR rather old, $O$(10\,000 yr), and place it in the late
Sedov-Taylor phase, or, more likely, in the radiative phase. If the SNR is in
the radiative phase, the ambient material swept up by the SNR should
be visible in CO or \ion{H}{I} data. Unfortunately, no high-resolution CO data
are publicly available at the moment, but there is evidence for a ring-like structure in
the \ion{H}{I} line emission survey of the Parkes telescope, as shown in
Fig.~\ref{fig3} (left). The structure is best visible in the
velocity range $-$13.79~km~s$^{-1}$ to $-$21.21~km~s$^{-1}$ corresponding to a
near/far kinematic distance of 1.6--2.3~kpc / 13.0--13.7~kpc (Fig.~\ref{fig3}, right).
The near distance is compatible with the pulsar distance (2.3~kpc) 
A rough
estimate of the mass of the \ion{H}{I} structure, extracted from the circular
region in Fig.~\ref{fig3} (left), is $\sim$6~$\times$~10$^3$~M$_{\sun}$. Assuming radial
symmetry, this corresponds to an original density of the swept-up mass
of a few protons~cm$^{-3}$, comparable to the density requirements
imposed by the observed $\gamma$-ray flux. The bright radio arc and
the VHE $\gamma$-ray emission spatially coincide with only one half of
the \ion{H}{I} shell-like structure. This morphology could arise because
of the additional dependence of the radio and $\gamma$-ray emission on the
target density, which is likely larger closer to the parent MC.

The \ion{H}{I} shell has a radius of $\sim$0.4\degr, which, assuming a distance of
$D$~$=$~2.3~kpc, corresponds to a physical radius of $\sim$16\,pc. Following
the approaches of \cite{Cioffi1988} and \cite{Truelove_1999} and further assuming an age of 17\,000
yr, a 10~M$_{\sun}$ progenitor star, and an energy release of
$10^{51}$~erg, the ambient density necessary to explain the size of
the \ion{H}{I} shell is $n$~$\sim$~0.7~cm$^{-3}$ and the resulting shock
velocity is $\sim$400~km~s$^{-1}$. Following
\cite{Ptuskin_2005}, the maximum proton energy attainable is then $O$(10~TeV),
likely too low to explain the observed TeV emission, which
extends up to 20~TeV.  The spectral energy distribution (SED) of
$\gamma$-rays produced in the interactions of mono-energetic protons (and subsequent
decay of pions) drops
sharply beyond roughly 15\% of the original proton energy, see
e.g. \citep{Kelner_2006}. Therefore, the parent proton population
giving rise to the observed VHE $\gamma$-ray emission should extend up
to about 100~TeV,
a limit which is -- as the example calculation
above illustrates -- increasingly difficult to explain as the age of
the system increases. Indeed, the $\gamma$-ray emitting SNR shells
which have been unambiguously identified so far, such
as \object{RX\,J1713.7$-$3946} \citep{Aharonian_2007_RXJ1713} and
\object{RX\,J0852.0$-$4622} \citep{Aharonian_2007_VelaJr}, are much younger
($\sim$2\,000~yr).

The aforementioned constraints are removed if the SNR expands 
first into a bubble blown by the progenitor star's wind into the ISM. 
Due to the low density inside the wind-blown
bubble, the velocity of the expanding shock is much higher than
anticipated and protons can be accelerated to very high
energies. When the shockfront reaches the outer boundary of the wind-blown
bubble, the high-energy protons are released to interact with
the dense environment outside of the bubble and produce VHE $\gamma$-rays
in the process. In this scenario, first proposed by \cite{Bock_2002}, 
the bright radio arc is created by the former
boundary of the wind-blown bubble which has been overtaken by the expanding
SNR shockfront.  The offset of the pulsar position from the center of
the radio arc does not hinder the association between the SNR and the
pulsar, since the progenitor star, whose wind has produced the bubble, can
have traversed the bubble's boundary during its evolution, before
it became a supernova. The constraints on the VHE $\gamma$-ray production
imposed by the large implied age of 17\,000 yr do not apply in this
case since the protons now interacting within the dense ambient medium
to produce $\gamma$-rays could have been accelerated in the past, when the
SNR shock velocity was still high.
However, this would require an extremely low
diffusion coefficient ($D_0~\approx~2~\times~10^{25}~\mathrm{cm}^2~\mathrm{s}^{-1}$),
similar to the case of the hadronic PWN scenario (see Sect.~4.1).

This discussion of a putative association between \hessj\
and the SNR \snr\ is based on the assumption that the SNR
and the pulsar \psr\ were created at the same time. If this
assumption proves to be wrong, then very little is known about the SNR.
The age estimate using a Sedov-Taylor model is about 5\,000 yr
\citep{McAdam_1993,Nicastro_1996}. The younger age would further ease
the proton acceleration to energies beyond 100~TeV.

To summarize, the radio emission from SNR \snr, which may originate from the interaction
of the SNR with an ambient MC, is partially coincident with \hessj, suggesting a plausible association
which could account for at least part of the VHE $\gamma$-ray emission observed. 
However, the putative associations between the SNR and the pulsar or between the SNR and the shell-like
structure discovered in \ion{H}{I} suggest that the SNR is in a 
later evolutionary stage than other SNRs previously-detected in the VHE regime. 

\subsection{Other Nearby Celestial Objects}
There are other celestial objects nearby, i.e. within the emission region of \hessj,
notably the LMXB 4U\,1705$-$440 \citep{Forman1978} and
the radio source \object{PMN\,J1708$-$4419} \citep{Wright_1994}.  The LMXB
is a well-studied type 1 burster \citep{Sztajno1985} located at
$\alpha_{2000}$~$=$~17$^{\mathrm{h}}$08$^{\mathrm{m}}$54.46$^{\mathrm{s}}$ and
$\delta_{2000}$~$=$~$-$44\degr6\arcmin7.35\arcsec\ \citep{DiSalvo2005}, i.e.\
it is offset from the centroid of the VHE emission by 0.25\degr. Considering this 
offset and the extended nature of the VHE $\gamma$-ray source, an association
is highly unlikely since an X-ray binary would appear point-like to H.E.S.S.
Theoretical predictions for
VHE $\gamma$-ray emission from LMXBs focus on those with relativistic jets (microquasars);
4U\,1705$-$440 does not exhibit jets.  Furthermore, 
no LMXBs have been detected in the VHE $\gamma$-ray regime, despite the
extensive coverage of the H.E.S.S.\ Galactic Plane Survey.

The radio source PMN\,J1708$-$4419 
is located at $\alpha_{2000}$~$=$~17$^{\mathrm{h}}$08$^{\mathrm{m}}$30$^{\mathrm{s}}$ and
$\delta_{2000}$~$=$~$-$44\degr19\arcmin07\arcsec\
\citep{Wright_1994}.
The local maximum in the radio contours at the center of Fig.~\ref{fig1} is largely 
due to this very bright point source, clearly visible in the 
330~MHz VLA radio image (see Fig.~\ref{figX}).
Although its position is compatible with the
centroid of the H.E.S.S.\ source, an association
between the two is unlikely given the spectrum of the radio
source.
Using data from the VLA (at 330~MHz and 1.4~GHz), 
Molonglo Galactic Plane Survey (MGPS; cataloged as \object{J\,170828$-$441823}
at 840~MHz, and Parkes-MIT-NRAO (PMN; cataloged as PMN\,J1708$-$4419 at 4.8~GHz), 
we derive a spectral index $\alpha = -0.81 \pm 0.08$ in the
radio domain, 
consistent with the value $\alpha = -0.9$ derived by \cite{Frail_1994} over a narrower range in frequency, 
from 330 to 840~MHz.
The steep spectral index suggests that PMN\,J1708$-$4419 is extragalactic, 
since Galactic point-like sources are typically compact \ion{H}{II} regions, 
for which the radio spectral index is positive;
therefore, it is unlikely to be associated with the extended emission of \hessj.
Upon a deeper inspection of this source using high-resolution unpublished
ATCA radio data (Dodson \& Golap, personal communication), this bright source can be further resolved into two sources.  
However, the spectral indices above were calculated on the basis of observing it as a single unresolved source,
because this is the way that the low-resolution radio surveys detected them.

In order to quantify the contribution any putative, unresolved point source could make to the flux observed from 
\hessj, one can compare the symmetric 2D Gaussian curve of a point source to that of the extended H.E.S.S.\
source.  This demonstrates that any single unresolved point source could not account for more than $\sim$6\% of the
total flux from \hessj.

\section{Summary}
H.E.S.S. observations of the $\gamma$-ray pulsar \psr\ have led to the detection
of an extended ($\sigma = 0.29\degr \pm 0.04\degr$)
source of VHE $\gamma$-ray emission, \hessj, in the Galactic plane.
Its energy spectrum is well-described by a power law with a photon index
$\Gamma$~$=$~2.0~$\pm$~0.1$_{\mathrm{stat}}$~$\pm$~$0.2_{\mathrm{sys}}$
and a normalization at 1~TeV
of $\phi_0~=~(4.2~\pm~0.8_{\mathrm{stat}}~\pm~1.0_{\mathrm{sys}})~\times 10^{-12}$~cm$^{-2}$~s$^{-1}$~TeV$^{-1}$.
The corresponding integral flux $F$(1--10~TeV)~$=$~$3.8~\times~10^{-12}$~ph~cm$^{-2}$~s$^{-1}$ is 
roughly 17\% of the Crab Nebula.
The possible associations of
\hessj\ with an offset, relic PWN of \psr\ and with the partial shell-type SNR
candidate \snr\ have been discussed using additional radio and \ion{H}{I} line emission data.
Given the extended nature of the TeV source and the limited statistics,
it is unclear if the emission is associated with the PWN,
located at the edge of the H.E.S.S.\ source, or with the SNR, in which the pulsar is thought to be embedded.
Based on energetics and a wealth of information at other wavelengths, neither interpretation can be excluded at this time; 
furthermore, the possibility remains that both sources contribute to the total observed VHE $\gamma$-ray emission.
High-spatial-resolution CO mapping of this region would improve our understanding of the molecular environment 
and might help
to identify a preferred MWL counterpart to \hessj.  Deeper exposure in the TeV regime would also provide
vital statistics and enable more detailed morphological and spectral studies.

\begin{acknowledgements}
  The support of the Namibian authorities and of the University of
  Namibia in facilitating the construction and operation of
  H.E.S.S. is gratefully acknowledged, as is the support by the German
  Ministry for Education and Research (BMBF), the Max Planck Society,
  the French Ministry for Research, the CNRS-IN2P3 and the
  Astroparticle Interdisciplinary Programme of the CNRS, the
  U.K. Science and Technology Facilities Council (STFC), the IPNP of
  the Charles University, the Polish Ministry of Science and Higher
  Education, the South African Department of Science and Technology
  and National Research Foundation, and by the University of
  Namibia. We appreciate the excellent work of the technical support
  staff in Berlin, Durham, Hamburg, Heidelberg, Palaiseau, Paris,
  Saclay, and in Namibia in the construction and operation of the
  equipment.
\\
 We would also like to thank R. Dodson and K. Golap for providing part
  of the radio data used in the analysis.
\end{acknowledgements}

\bibliographystyle{aa} 
\bibliography{15381} 

\appendix
\section{Comparison with H.E.S.S.\ 2003 dataset}

\subsection{Recalculation of upper limits using the 2003 dataset}

In its commissioning phase, 
the H.E.S.S.\ IACT observed the region around the energetic \psr\ between April and July 2003
\citep{Aharonian_2005_1706}.
No evidence for statistically-significant VHE $\gamma$-ray emission was found at the 
pulsar position nor from a region encompassing the partial shell-type SNR \snr.
Upper limits (ULs) to the integral flux were published in \citep{Aharonian_2005_1706}.
The integral flux now measured by H.E.S.S.\ (see Sect.~3) is not
compatible with those originally-published ULs, a discrepancy which 
motivated a re-analysis of the 2003 H.E.S.S.\ dataset for this region, using the current
H.E.S.S.\ software.

Although H.E.S.S.\ is currently an array of four IACTs, it was operating as a 
two-telescope array from February to December 2003.  
The 2003 observations yielded a dataset with a livetime of 14.3~h, an average zenith angle of $\sim$26\degr, and an
energy threshold, estimated from Monte Carlo simulations, of $\sim$350~GeV.  For $\gamma$-hadron separation,
\emph{standard cuts} were used, which require a minimum of 80 p.e.\ to be recorded per shower image.

Integral flux ULs were calculated from three \emph{a priori} defined circular regions: a \emph{Standard} point-like
($\theta = 0.14\degr$) region centered at the 
position of \psr, a \emph{CANGAROO}-like region ($\theta = 0.22\degr$) also centered at the pulsar position, and a
\emph{Radio arc} region centered at the
apparent center of the SNR \snr\ ($\alpha_{2000}$~$=$~17$^{\mathrm{h}}$08$^{\mathrm{m}}$,
$\delta_{2000}$~$=$~$-$44\degr16\arcmin48\arcsec; as defined
in \cite{Aharonian_2005_1706}), with a radius $\theta = 0.60\degr$ in order to completely enclose the complex radio 
structure.  The \emph{CANGAROO} region is disregarded
for the remainder of this appendix,
because its sole purpose was to compare the H.E.S.S.\ result with the original CANGAROO-I detection,
which has since been rescinded \citep{Yoshikoshi2009}, and
focus primarily on the \emph{Radio arc} region (equivalent to Region B; see Sect.~3), 
which is very similar to the region from which extended VHE $\gamma$-rays are now detected (Region C; see Sect.~3).

\begin{figure}[t!]
 \begin{center}
    \includegraphics[width=0.49\textwidth]{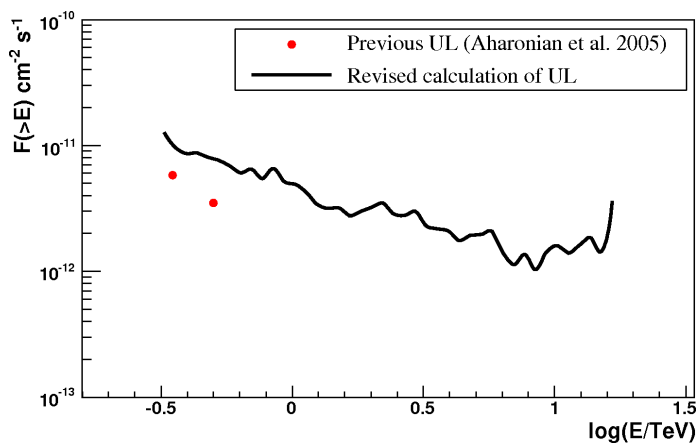}
    \caption{Integral flux upper limits (ULs) from the \emph{Radio arc} region encompassing both \psr\ and SNR \snr. 
      The red points represent the previously-published integral flux ULs from \cite{Aharonian_2005_1706} using the 2003 
      H.E.S.S.\ dataset.  The black solid line is based on a revised calculation of the UL using the same dataset.}
    \label{figA1}
 \end{center}
\end{figure}

Background subtraction was performed using the Ring Background Method \citep{Berge_2007}
in both the original and revised analysis.
The exact inner and outer ring radii, $r_{\mathrm{inner}}$ and $r_{\mathrm{outer}}$  
respectively, used in the original analysis were not documented;
however, the inner ring radius is typically chosen to be slightly larger than the on-source (ON) region
(radius $\theta = 0.60\degr$) and
the normalization factor $\alpha = 1/7$ (the ratio of the ON to off-source (OFF) area).  
Therefore, for the re-analysis, the inner ring radius is chosen to be 0.65\degr, which, given $\alpha$,
leads to $r_{\mathrm{outer}} = 1.35\degr$.  The number of events in the ON and OFF regions, $N_{\mathrm{ON}}$ and
$N_{\mathrm{OFF}}$, respectively, 
is found to match those given in \cite{Aharonian_2005_1706} to within 8\% (see Table~\ref{Table}),
demonstrating that the ring parameters adopted in the re-analysis are approximately equal to
those in the original analysis. No exclusion region was placed on the now known-to-exist source, \hessj, 
i.e.\ the source is not excluded from OFF regions.  In practice this has a negligible effect, since there is 
little $\gamma$-ray emission from HESS\,J1708$-$443 beyond $r_{\mathrm{inner}} = 0.65\degr$ from its centroid.

The UL (99\% confidence level using \citet{Feldman_1998}) 
on the integral flux from the \emph{Radio arc} region (Region B)
was originally found to be $F$($>$~0.35~TeV)~$<$~$5.8 \times 10^{-12}$~ph~cm$^{-2}$~s$^{-1}$,
equivalent to $\sim$5\% Crab, 
assuming the spectrum is described by a power law with a spectral index $\Gamma = 2.5$ (\emph{Method A} in
\cite{Aharonian_2005_1706}).
An alternative UL, $F$($>$~0.50~TeV)~$<$~$3.5 \times 10^{-12}$~ph~cm$^{-2}$~s$^{-1}$, also equivalent to $\sim$5\% Crab,
was calculated using a method (\emph{Method B}, described in detail in
\citet{Aharonian_2005_1706}) which made no assumptions concerning the source spectrum.
These ULs are shown in Fig.~\ref{figA1}, where they are compared to the revised calculation (using \emph{Method A})
of the integral flux UL, plotted as a function of threshold energy $E$.
The revised UL is clearly higher (less stringent) than the one published in \cite{Aharonian_2005_1706}.  For example,
the integral flux above 0.35~TeV is $F$($>$~0.35~TeV)~$<$~$9.7 \times 10^{-12}$~ph~cm$^{-2}$~s$^{-1}$, equivalent
to 9\%~Crab and above 0.50~TeV is $F$($>$~0.50~TeV)~$<$~$7.6 \times 10^{-12}$~ph~cm$^{-2}$~s$^{-1}$, equivalent
to 12\%~Crab, again assuming $\Gamma = 2.5$.
See Table~\ref{Table} for a summary and comparison of the event statistics and other 
analysis parameters from both analyses.  

\begin{table}[]
  \begin{center}
    \begin{tabular}{c c c}
      \hline\hline
                                                   & \cite{Aharonian_2005_1706}  & Re-analysis             \\
      \hline
      $F(> 0.35 \mathrm{TeV})$ (cm$^{-2}$~s$^{-1}$) & $< 5.8 \times 10^{-12}$      & $< 9.7 \times 10^{-12}$  \\
      $N_{\mathrm{ON}}$                              & 4746                        & 5095                    \\ 
      $N_{\mathrm{OFF}}$                             & 13688                       & 14730                   \\ 
      $\alpha$                                     & 0.346                       & 0.343                   \\ 
      Excess                                       & 11                          & 38                      \\ 
      Significance                                 & 0.1 $\sigma$                & 0.5 $\sigma$            \\ 
      $r_{\mathrm{inner}}$                           & $> 0.60\degr$               & 0.65\degr               \\
      $r_{\mathrm{outer}}$                           & unknown                     & 1.35\degr               \\ 
      \hline
    \end{tabular}
    \caption{Event statistics and background parameters for the analyses of the \emph{Radio arc} region around \psr\ and
      SNR \snr. Row 1 gives the integral flux upper limits (99\% confidence level) from both analyses.
      The number of events $N$ in the circular (radius $\theta = 0.6\degr$) on-source (ON) and ring-shaped off-source
      (OFF) regions are given in rows 2 and 3,
      the normalization factor $\alpha$ (the ratio of ON to OFF area) in row 4, excesses and significances (according to 
      \cite{LiMa_1983}) in rows 5 and 6,
      and the ring parameters in rows 7 and 8.     
      The statistics and upper limits presented here were obtained using the 2003 H.E.S.S.\ dataset only.}
      \label{Table}
  \end{center}
\end{table}

The use of two-telescope data 
resulted in a lower sensitivity at the time but would not have had any negative impact on the original
determination of ULs from the vicinity of \psr.
H.E.S.S.\ currently uses a stereo trigger implemented at the \emph{hardware} level to select
extended air showers (EASs) simultaneously detected by at least two telescopes \citep{Funk2004}.
However, from February to July 2003, when the original observations of \psr\ were carried out, 
it used an off-line triggering mode, since the central hardware trigger had not yet been installed.
In \emph{software} stereo mode, each recorded EAS receives a time stamp via a GPS (Global Positioning System) clock.
The time stamps are then used in the offline data analysis to identify EASs which were observed in coincidence by
the two telescopes.  The use of a software stereo trigger, while not as efficient as the hardware stereo trigger
currently in use, is not expected to have contributed significantly to the discrepancy between the original ULs and
the new results.

After investigating various possible reasons for the discrepancy, 
it remains unknown why the previously-determined ULs were so low, leaving human error or
undocumented changes in the analysis software used at the time as possible explanations.
It is important to note that many other published results based on data taken during H.E.S.S.'s
commissioning phase have been subsequently confirmed by the full four-telescope array with a hardware trigger,
e.g. observations of RX\,J1713$-$3946 \citep{Aharonian_2007_RXJ1713} and \object{Sgr\,A$^{*}$} \citep{SgrA}.

\subsection{Compatibility between detected flux and 2003 upper limits}
The upper limits calculated in the previous section cannot be directly compared to the new H.E.S.S. results, 
based on the 2007 dataset (presented in Sect.~3), because they assume a spectral
index $\Gamma = 2.5$ and a low energy threshold. The new VHE $\gamma$-ray source, \hessj, has a 
much harder spectral index $\Gamma$~$=$~2.0~$\pm$~0.1$_{\mathrm{stat}}$~$\pm$~$0.2_{\mathrm{sys}}$.  
Furthermore, the minimum energy threshold of the H.E.S.S.\ array has increased due to 
the reflectivity of the IACT mirrors diminishing from 2003 to 2007, 
which reduces the array's ability to detect faint EASs initiated by lower-energy $\gamma$-rays.

A re-analysis of the 2003 dataset, using the current H.E.S.S.\ software
and assuming $\Gamma = 2.0$, yields a flux upper limit (99\% confidence level)
$F$($>$~0.6~TeV)~$<$~$6.3 \times 10^{-12}$~ph~cm$^{-2}$~s$^{-1}$, equivalent to $\sim$13\% Crab, for Region B.
Analysis of the 2007 dataset shows a statistically-significant signal from Region B, 
$F$($>$~0.6~TeV)~$\approx$~$6.5 \times 10^{-12}$~ph~cm$^{-2}$~s$^{-1}$, equivalent to $\sim$13\% Crab.  These
two flux values are statistically compatible,
given the typical uncertainties in the measured flux normalization ($\pm \sim$20\%) and spectral index ($\pm \sim$0.2).

Furthermore, the new analysis results for HESS\,J1708$-$443 (Region C), based on the 2007 dataset,
have been confirmed using an independent data calibration and analysis chain, and 
the cross-check analysis is also compatible with the presented results.  

\listofobjects

\end{document}